\documentclass[referee]{aa}

\begin{document}

\title{Electromagnetic radiation, motion of a particle and energy-mass relation}
\author{J. Kla\v{c}ka}
\institute{Faculty of Mathematics,
Physics and Informatics, Comenius University \\
Mlynsk\'{a} dolina, 842 48 Bratislava, Slovak Republic \\
e-mail: klacka@fmph.uniba.sk}

\date{}

\abstract{
Equation of motion of an uncharged arbitrarily shaped dust particle under the
effects of (stellar) electromagnetic radiation and thermal emission is derived.
The resulting relativistically covariant equation of motion is expressed
in terms of standard optical parameters. Relations between energy
and mass of the incoming and outgoing radiation are obtained, together with
relations between radiation energy and mass of the particle.
The role of the diffraction nicely fits the relativistic formulation of the
momentum of the outgoing radiation.

Covariant formulations yield several simple consequences.
It is shown that the frequently used statement
''energy equals $\gamma$ $\times$ mass $\times$ $c^{2}$'' or
''energy equals mass $\times$ $c^{2}$'' are not correct, in general;
"rest energy equals mass $\times$ $c^{2}$" holds.
The inequality 0 $<$ $\bar{Q}'_{pr, 1} / \bar{Q}'_{ext}$ $<$ 2 is a simple
relativistic consequence for the Poynting-Robertson (P-R) effect
($\bar{Q}'_{ext}$ and $\bar{Q}'_{pr, 1}$ are dimensionless efficiency factors
for the extinction and radial direction of the radiation pressure, integrated
over stellar spectrum).
The condition for the P-R effect is
$\vec{p}'_{o}$ $=$ ( 1 $-$ $\bar{Q}'_{pr, 1} / \bar{Q}'_{ext}$ ) $\vec{p}'_{i}$,
where $\vec{p}'_{i}$ and $\vec{p}'_{o}$ are incoming and outgoing radiation
momenta (per unit time) measured in the proper frame of reference of the
particle. The case of "perfectly absorbing spherical dust particle",
within geometrical optics approximation, corresponds
to the condition $\vec{p}'_{o}$ $=$ 0.5 $\vec{p}'_{i}$.
The obtained results enable understanding of the physics of the P-R effect
and evaluate the published explanations.
As for arbitrarily shaped dust particle, the general condition
0 $<$ $\bar{C}'_{pr, 1}$ $/$ $\bar{C}'_{ext}$  $<$ 2 $/$
($1 ~+~ \sum_{j=2}^{3}	\bar{C}'_{pr, j} / \bar{C}'_{pr, 1}$) holds for
cross sections of extinction and radiation pressure components, if
thermal emission force is of negligible importance. The condition
can add a new information to the results obtained from observations,
measurements and numerical calculations of the optical properties
of the particle.

}

\authorrunning{J. Kla\v{c}ka}
\titlerunning{Electromagnetic radiation, motion of a particle}
\maketitle

\section{Introduction}
The Poynting-Robertson effect is used in modelling
orbital evolution of dust grains under the action of electromagnetic
radiation (of the central star), for many decades (e. g., Poynting 1903,
Robertson 1937, Wyatt and Whipple 1950, Dohnanyi 1978,
Kapi\v{s}insk\'{y} 1984, Jackson and Zook 1989,
Leinert and Gr\H{u}n 1990, Gustafson 1994,  Dermott {\it et al.} 1994,
Reach {\it et al.} 1995, Murray and Dermott 1999,
Woolfson 2000, Danby 2003,
Quinn 2005, Harwit 2006, Gr\H{u}n 2007, Sykes 2007, Kr\"{u}gel 2008).
It is assumed that particle with spherically distributed material can be used
as a correct approximation to reality.
However, real nonspherical particles interact with electromagnetic radiation in
a complicated manner and particles of various optical properties
exist (e. g., Mishchenko {\it et al.} 2002). Thus,
it is essential to have an equation of motion sufficiently
general to cover a wide range of optical parameters, not just the
limited cases standardly investigated. The first presentations of the equation
of motion for arbitrarily shaped particle were given by Kla\v{c}ka (1994),
Kla\v{c}ka and Kocifaj (1994), Kla\v{c}ka (2000), Kocifaj {\it et al.} (2000),
Kla\v{c}ka and Kocifaj (2001). Relativistically covariant derivation can
be found in Kla\v{c}ka (2000) and Kla\v{c}ka (2004), where also thermal
emission of the particle is taken into account -- Mishchenko (2001) has
formulated the radiation pressure on arbitrarily shaped particles arising
from an anisotropy of thermal emission. Krauss and Wurm (2004) have published
the first experimental evidence that nonspherical dust grains behave
in a different way than the spherical ones.

We derive equation of motion for arbitrarily shaped particle in a
physically correct way and relativistic covariant formulations are presented.
A new form for the validity of the Poynting-Robertson (P-R) effect is found:
the P-R effect holds only when the total momentum of the
outgoing radiation per unit time is colinear with the incident
radiation, in the proper/rest frame of reference of the particle, and, a new
physical condition for the P-R effect is found, as for
the dimensionless efficiency factors for radiation pressure and extinction
(optical properties) of the particle. Moreover,
relations between energy and mass of the incoming and outgoing
radiation are presented, both for arbitrarily shaped particle and also, as a
special case, for spherical particle for which the P-R effect
holds. The relations between energy and mass for the incoming and
outgoing radiation, together with relations between radiative energy and mass
of the particle, present a practical contribution to the discussion
on relativistic energy-mass relation (see, e. g.,
Ugarov 1969, Secs. 57-58, pp. 246-252; Okun 1989a, 1989b;
Taylor and Wheeler 1992, Sec. 8.4, pp. 228-233; Einstein 1999).
The relations are consistent with the consideration of the diffracted
light within geometrical optics approximation.
The results of the paper, including detail discussion also with respect
to the statements presented in the original papers by Poynting (1903) and
Robertson (1937), enable to understand physics of the
P-R effect, which is frequently used in studies of orbital
evolution of cosmic dust particles. As for the
arbitrarily shaped dust particles, a condition yielding some
information on cross sections, unknown from experiments,
observations or theoretical/numerical solutions is obtained as a simple
relativistic consequence of the covariant form of the four-momentum
of the outgoing radiation.

We begin by reviewing in Sec. 2 and 3 the basic physical processes in proper
and stationary frames, for an arbitrarily shaped particle.
The physical equations use cross sections for all
relevant phenomena (extinction, scattering, absorption, radiation
pressure components) and the equation of motion of the particle for the case of
the irradiated particle is discussed at the end of Sec. 3 (Eqs. 40-41).
The following Sec. 4
deals with the covariant formulation for the incoming and outgoing radiation,
which are used in Sec. 5 for finding energy-mass relation for the radiation.
Sec. 6 discusses the relation between the change of the particle's mass,
separately for the incoming and outgoing radiation. Application of the
results from the previous sections are discussed, for special particle shapes,
in Sec. 7: (i) it is shown that a simple relations holds for dimensionless
efficiency factors for radiation pressure and extinction for the case
of spherical particle (Eq. 79), including presentation of the value
for perfectly absorbing (or perfectly reflecting) sphere (Eq. 104) which
differs from the conventional statement coming back to Poynting and Robertson;
(ii) an understanding of the action of the radiation
on motion of spherical particles is discussed in detail in Sec. 7.4;
(iii) an analogy between mechanics and electromagnetism for the case
of planar surface is discussed in Sec. 7.5. Results of Secs. 2-7 are used
in Sec. 8, which presents considerations on the equivalence principle of mass
and energy for absorbing planar surfaces. A special Sec. 9 derives a simple
condition for cross sections for radiation pressure components and extinction,
using results of Sec. 5. The most important results are shortly summarized
in Sec. 10.

\section{Proper reference frame of the particle -- stationary particle}
The term ``stationary particle'' will denote a particle which does
not move in a given inertial frame of reference.
Primed quantities will denote quantities measured in the
proper reference frame of the particle -- rest frame of the particle.

The flux density of photons scattered into an elementary solid
angle $d \Omega ' = \sin \theta ' ~ d \theta ' ~ d \phi '$ is
proportional to  $p' ( \theta ', \phi ') ~ d \Omega '$, where $p'
( \theta ', \phi ')$ is the ``phase function''. The phase function
depends on orientation of the particle with respect to the
direction of the incident radiation and on the particle
characteristics; angles $\theta '$, $\phi '$ correspond to the
direction (and orientation) of travel of the scattered radiation,
$\theta '$ is the polar angle which vanishes for propagation along
the unit vector $\vec{e}_{1} '$ of the incident radiation. The
phase function fulfills the normalisation condition
\begin{equation}\label{1}
\int_{4 \pi} p' ( \theta ', \phi ')~ d \Omega ' = 1 ~.
\end{equation}

The momentum of the incident beam of photons which is lost in the
process of interaction with the particle is proportional to the
cross section $C'_{ext}$ (extinction). The part proportional to
$C'_{abs}$ (absorption) is emitted in the form of thermal
radiation and the part proportional to
$C'_{ext} ~-~ C'_{abs} = C'_{sca}$ is scattered.
The differential scattering cross section
$dC'_{sca}/d \Omega '$ $\equiv$ $C'_{sca} ~p'(\theta ', \phi ' )$
depends on the polarization state of the incident light as well as
on the incidence and scattering directions (e. g., Mishchenko {\it
et al.} 2002).

The momentum (per unit time) of the scattered photons into an elementary
solid angle $d \Omega '$ is
\begin{equation}\label{2}
d \vec{p'}_{sca} = \frac{1}{c} ~ S' ~ C'_{sca} ~
       p' ( \theta ', \phi ')~ \vec{K'} ~ d \Omega ' ~,
\end{equation}
where the unit vector in the direction of scattering is
\begin{equation}\label{3}
\vec{K'} = \cos \theta '~\vec{e}_{1} ' ~+~
     \sin \theta ' ~ \cos \phi ' ~ \vec{e}_{2} ' ~+~
     \sin \theta ' ~ \sin \phi ' ~ \vec{e}_{3} ' ~.
\end{equation}
$S'$ is the flux density of radiation energy
(energy flow through unit area perpendicular to the ray per unit time).
The system of unit vectors
used on the RHS of the last equation forms an orthogonal basis.
The total momentum (per unit time) of the scattered photons is
\begin{equation}\label{4}
\vec{p'}_{sca} = \frac{1}{c} ~ S' ~ C'_{sca} ~ \int_{4 \pi} ~
     p' ( \theta ', \phi ')~ \vec{K'} ~ d \Omega ' ~.
\end{equation}

The momentum (per unit time) obtained by the particle due to the interaction
with radiation -- radiation force acting on the particle -- is
\begin{equation}\label{5}
\frac{d~ \vec{p'}}{d~ t'} = \frac{1}{c} ~ S' ~ \left \{
      C'_{ext} ~\vec{e}_{1} '
      ~-~ C'_{sca} ~ \int_{4 \pi} ~
      p' ( \theta ', \phi ')~ \vec{K'} ~
      d \Omega ' \right \} ~+~ \vec{F}'_{e} ( T' ) ~,
\end{equation}
where the emission component of the radiation force acting on the particle
of absolute temperature $T'$ is (Mishchenko {\it et al.} 2002, pp. 63-66)
\begin{equation}\label{6}
\vec{F}'_{e} ( T' ) = -~ \frac{1}{c} ~ \int_{0}^{\infty} ~ d \omega' ~
      \int_{4 \pi} ~ \hat{\vec{r}}' ~
      K'_{e} \left ( \hat{\vec{r}}', T', \omega ' \right ) ~
      d \hat{\vec{r}}' ~.
\end{equation}
The unit vector $\hat{\vec{r}}' = \vec{r}' / r'$ is given by position vector
$\vec{r}'$ of the observation point with origin inside the particle
(the emitted radiation in the far-field zone of the particle propagates
in the radial direction, i. e., along the unit vector $\hat{\vec{r}}'$),
$\omega '$ is (angular) frequency of radiation,
\begin{equation}\label{7}
K'_{e} \left ( \hat{\vec{r}}', T', \omega ' \right ) =
       I'_{b} \left ( T', \omega ' \right )  \left [
       K'_{11} \left ( \hat{\vec{r}}', \omega ' \right ) -
       \int_{4 \pi}
       Z'_{11} \left ( \hat{\vec{r}}', \hat{\vec{r}}'', \omega ' \right )
      d \hat{\vec{r}}'' \right ] ~,
\end{equation}
where $K'_{11}$ is the (1,1) element of the particle extinction matrix,
$Z'_{11}$ is the (1,1) element of the phase matrix and
the Planck blackbody energy distribution is given by the well-known relation
\begin{equation}\label{8}
I'_{b} \left ( T', \omega ' \right ) =
       \frac{\hbar ~\omega'^{3}}{4~ \pi ^{3} ~c^{2}}
       \left \{ \exp \left ( \frac{\hbar ~\omega'}{k~T'} \right )
       ~-~ 1 \right \}^{-1} ~.
\end{equation}
Thermal emission has to be included in the total interaction of the particle
with electromagnetic radiation:
if the particle's absolute temperature is above zero, it can emit as well
as scatter and absorb electromagnetic radiation. The particle is assumed
to be isothermal, in Eqs. (6)-(8).

Equation (5) can be rewritten to the form
\begin{eqnarray}\label{9}
\frac{d ~\vec{p'}}{d~ \tau} &=& \frac{1}{c} ~ S' ~ \left \{ \left [
       C'_{ext} ~-~ < \cos \theta'> ~ C'_{sca} \right ] ~
       \vec{e}_{1} ' ~+~
       \right .
\nonumber \\
& &  \left .
       \left [ ~-~ < \sin \theta' ~ \cos \phi ' > ~
       C'_{sca} \right ] ~ \vec{e}_{2} ' ~+~
       \right .
\nonumber \\
& &  \left .
       \left [ ~-~ < \sin \theta' ~ \sin \phi ' > ~ C'_{sca}
       \right ] ~ \vec{e}_{3} ' \right \} ~+~
       \sum_{j=1}^{3} ~F'_{e, j} ~ \vec{e}'_{j} ~,
\end{eqnarray}
where $< x' > \equiv  \int_{4 \pi} ~ x' ~p' ( \theta ', \phi ') ~
d \Omega '$ and $F'_{e, j} \equiv \vec{F}'_{e} ( T' ) \cdot \vec{e}'_{j}$.
As for the energy, we assume that it is conserved: the energy (per
unit time) of the incoming radiation $E'_{i}$, equals to the
energy (per unit time) of the outgoing radiation (after
interaction with the particle) $E'_{o}$. We will use the fact that
time $t' = \tau$, where $\tau$ is proper time.

Summarizing important equations, we can write them in a short form
\begin{equation}\label{10}
\frac{d~ \vec{p}'}{d~ \tau} = \sum_{j=1}^{3} \left ( \frac{S'}{c} ~
     C'_{pr, j} ~+~ F'_{e, j}  \right ) ~\vec{e}_{j} ' ~;~~
\frac{d ~E'}{d~ \tau} = 0 ~,
\end{equation}
where $C'_{pr, 1} \equiv C'_{ext} ~-~ < \cos \theta'> ~ C'_{sca}$,
$C'_{pr, 2} \equiv  -~ < \sin \theta' ~ \cos \phi ' > ~ C'_{sca}$,
$C'_{pr, 3} \equiv -~ < \sin \theta' ~ \sin \phi ' > ~ C'_{sca}$
are cross sections for radiation pressure.
We have added an assumption of equilibrium state when the particle's
mass does not change.

\subsection{Summary of the important equations}
Using the text concerning energy below Eq. (9) and the last Eq. (10), we
may describe the total process of interaction in the form of the
following equations (energies and momenta per unit time):
\begin{eqnarray}\label{11}
E_{o} ' &=& E_{i} ' = S'~C'_{ext}  ~,
\nonumber \\
\vec{p}_{o} ' &=& \left ( 1 ~-~ \frac{C'_{pr, 1}}{C'_{ext}} \right ) ~
		  \vec{p}_{i} ' ~-~ \left (
      \frac{C'_{pr, 2}}{C'_{ext}} ~ \vec{e}_{2} ' ~+~
      \frac{C'_{pr, 3}}{C'_{ext}} ~ \vec{e}_{3} ' \right ) ~ \frac{E'_{i}}{c} ~-~
       \sum_{j=1}^{3} ~F'_{e, j} ~ \vec{e}'_{j} ~,
\nonumber \\
\vec{p}_{i} ' &=& \frac{E_{i} '}{c} ~ \vec{e}_{1} ' ~.
\end{eqnarray}
The index $"i"$ represents the incoming (incident) radiation,
beam of photons, the index $"o"$ represents the outgoing radiation.

The changes of energy and momentum of the particle due to the interaction
with electromagnetic radiation are
\begin{eqnarray}\label{12}
\frac{d ~E'}{d~ \tau} &=& E_{i} ' ~-~ E_{o} ' = 0 ~,
\nonumber \\
\frac{d ~\vec{p'}}{d~ \tau} &=& \vec{p}_{i} ' ~-~ \vec{p}_{o} ' ~.
\end{eqnarray}

\section{Stationary frame of reference}
By the term ``stationary frame of reference''
(laboratory frame) we mean a frame of reference
in which particle moves with a velocity vector $\vec{v} = \vec{v} (t)$.
The physical quantities measured in the stationary frame of reference
will be denoted by unprimed symbols.

Our aim is to derive equation of motion for the particle in the
stationary frame of reference. We will use the fact that we know
this equation in the proper frame of reference -- see Eqs. (11) and
(12).

If we have a four-vector $A^{\mu} = ( A^{0}, \vec{A} )$, where
$A^{0}$ is its time component and $\vec{A}$ is its spatial component,
generalized special Lorentz transformation yields
\begin{eqnarray}\label{13}
A^{' 0}  &=& \gamma ~ ( A^{0} ~-~ \vec{v} \cdot \vec{A} / c ) ~,
\nonumber \\
\vec{A} ' &=& \vec{A} ~+~ [ ( \gamma ~-~ 1 ) ~ \vec{v} \cdot \vec{A}  /
      \vec{v} ^{2} ~-~ \gamma ~ A^{0} / c ] ~ \vec{v}  ~,
\end{eqnarray}
with inverse
\begin{eqnarray}\label{14}
A^{0}  &=& \gamma ~ ( A^{' 0} ~+~ \vec{v} \cdot \vec{A} ' / c ) ~,
\nonumber \\
\vec{A}  &=& \vec{A} ' ~+~ [ ( \gamma ~-~ 1 ) ~ \vec{v} \cdot \vec{A} ' /
      \vec{v} ^{2} ~+~ \gamma ~ A^{' 0} / c ] ~ \vec{v}  ~,
\end{eqnarray}
where
$\gamma = 1 / \sqrt{1~-~\vec{v} ^{2} / c ^{2} }$ ~.

As for four-vectors we immediately introduce four-momentum:
\begin{equation}\label{15}
p^{\mu} = ( p^{0}, \vec{p} ) \equiv ( E / c, \vec{p} ) ~.
\end{equation}

\subsection{Incoming radiation}
Applying Eqs. (14) and (15) to quantity $( E_{i} ' / c, \vec{p}_{i} ')$
(four-momentum per unit time -- proper time is a scalar quantity) and
taking into account also Eqs. (11), we can write
\begin{eqnarray}\label{16}
E_{i}  &=& E_{i} ' ~ \gamma ~ ( 1 ~+~ \vec{v} \cdot \vec{e}_{1} ' / c ) ~,
\nonumber \\
\vec{p}_{i}  &=& \frac{E_{i} '}{c}  ~ \left \{ \vec{e}_{1} ' ~+~
     \left [ \left ( \gamma ~-~ 1 \right ) ~
     \vec{v} \cdot \vec{e}_{1} '  /
     \vec{v} ^{2} ~+~ \gamma / c \right ] ~ \vec{v} \right \} ~.
\end{eqnarray}

Using the fact that $p^{\mu} = ( h ~\nu / c, h ~\nu ~ \vec{e}_{1} / c)$
for photons, we have
\begin{eqnarray}\label{17}
\nu ' &=& \nu  ~w_{1} ~,
\nonumber \\
\vec{e}_{1} ' &=& \frac{1}{w_{1}}  ~ \left \{ \vec{e}_{1} ~+~
     \left [ \left ( \gamma ~-~ 1 \right ) ~
     \vec{v} \cdot \vec{e}_{1}	/
     \vec{v} ^{2} ~-~ \gamma / c \right ] ~ \vec{v} \right \} ~,
\end{eqnarray}
where
\begin{equation}\label{18}
w_{1} \equiv \gamma ~ ( 1 ~-~ \vec{v} \cdot \vec{e}_{1} / c ) ~.
\end{equation}

Inserting the second of Eqs. (17) into Eq. (16), one obtains
\begin{eqnarray}\label{19}
E_{i} &=& ( 1 / w_{1} ) ~ E_{i} ' ~,
\nonumber \\
\vec{p}_{i} &=& ( 1 / w_{1} ) ~ ( E_{i} ' / c ) ~ \vec{e}_{1} ~.
\end{eqnarray}
We have four-vector $p_{i}^{\mu} = ( E_{i} / c, \vec{p}_{i} ) = (
1, \vec{e}_{1} ) E_{i} / c$  $= ( 1 / w_{1}, \vec{e}_{1} / w_{1} )$
$\times$ $w_{1} ~ E_{i} / c$ $\equiv$ $b_{1}^{\mu}~w_{1} ~ E_{i} / c$.

For monochromatic radiation the flux density of radiation energy
becomes
\begin{equation}\label{20}
S' = n' ~h~ \nu ' ~ c ~; ~~~ S = n ~h~ \nu  ~ c ~,
\end{equation}
where $n$ and $n'$ are concentrations of photons (photon number
densities) in the corresponding reference frames. We also have
continuity equation
\begin{equation}\label{21}
\partial _{\mu} ~ j^{\mu} = 0 ~, ~~~j^{\mu} = ( c~n, c~n~ \vec{e}_{1} ) ~,
\end{equation}
with current density $j^{\mu}$. Application of Eq. (13) then
yields
\begin{equation}\label{22}
n' = w_{1} ~ n ~.
\end{equation}
Using Eqs. (17), (20) and (22) we finally obtain
\begin{equation}\label{23}
S' = w_{1}^{2} ~ S ~.
\end{equation}
Eqs. (11), (19) and (23) then together give $E_{i} = w_{1} ~ S
~C'_{ext}$, $\vec{p}_{i} = w_{1} ~S ~C'_{ext} ~\vec{e}_{1} / c$.

\subsection{Outgoing radiation}
The situation is analogous to that of the preceding subsection. It
is only a little more algebraically complicated, since radiation
may also spread out in directions given by unit vectors
$\vec{e}_{2}$, $\vec{e}_{3}$. We need transformations
$\vec{e}_{j} '$ $\rightarrow$ $\vec{e}_{j}$, $j$ $=$ 2, 3. The vectors
$\vec{e}_{2} '$ and $\vec{e}_{3} '$ can be used to describe directions
of propagation of radiation scattered by the particle.
Thus,  aberration of light also exists for each of these unit
vectors. The relations between $\vec{e}_{2} '$ and $\vec{e}_{2}$,
$\vec{e}_{3} '$ and $\vec{e}_{3}$, are analogous to that presented
by the second of Eq. (17):
\begin{equation}\label{24}
\vec{e}_{j} ' = \frac{1}{w_{j}}  ~ \left \{ \vec{e}_{j} ~+~
     \left [ \left ( \gamma ~-~ 1 \right ) ~
     \vec{v} \cdot \vec{e}_{j}	/
     \vec{v} ^{2} ~-~ \gamma / c \right ] ~ \vec{v} \right \} ~,
     ~~ j = 1, 2, 3 ~,
\end{equation}
where
\begin{equation}\label{25}
w_{j} \equiv \gamma ~ ( 1 ~-~ \vec{v} \cdot \vec{e}_{j} / c )
      ~, ~~ j = 1, 2, 3 ~.
\end{equation}
It is worth mentioning that vectors
$\left \{ \vec{e}_{j} ' ; j = 1, 2, 3 \right \}$ form an orthonormal
set of vectors, and, unit vectors
$\left \{ \vec{e}_{j} ; j = 1, 2, 3 \right \}$ are not orthogonal unit
vectors.

Applying Eqs. (14) and (15) to the quantity $( E_{o} ' / c,
\vec{p}_{o} ')$ (four-momentum per unit time -- proper time is a
scalar quantity), we can write
\begin{eqnarray}\label{26}
E_{o}  &=&  \gamma ~ ( E_{o} ' ~+~ \vec{v} \cdot \vec{p}_{o} ' ) ~,
\nonumber \\
\vec{p}_{o}  &=& \vec{p}_{o} ' ~+~
     \left [ \left ( \gamma ~-~ 1 \right ) ~
     \frac{\vec{v} \cdot \vec{p}_{o} '}{\vec{v} ^{2}} ~+~
     \gamma ~ \frac{E_{o} '}{c^{2}} ~\right ] ~ \vec{v}  ~.
\end{eqnarray}
Using also $\vec{p}_{i} '= E_{i} ' ~\vec{e}_{1} '/ c$ and Eqs. (11), (24),
(26),
\begin{eqnarray}\label{27}
E_{o}  &=& \frac{C'_{pr, 1}}{C'_{ext}}
	   ~ w_{1} ~ E_{i} ~ \gamma  ~+~ \left ( 1 ~-~
	   \frac{C'_{pr, 1}}{C'_{ext}} \right ) ~ E_{i}
\nonumber \\
& &  +~  w_{1} ~ E_{i} ~ \left ( \frac{C'_{pr, 2}}{C'_{ext}}  ~+~
			 \frac{C'_{pr, 3}}{C'_{ext}}  \right ) ~ \gamma  ~-~
       w_{1} ~ E_{i} ~ \left ( \frac{C'_{pr, 2}}{C'_{ext}} ~ \frac{1}{w_{2}} ~+~
       \frac{C'_{pr, 3}}{C'_{ext}} ~ \frac{1}{w_{3}} \right )
\nonumber \\
& & -~ \sum_{j=1}^{3} F'_{e, j} ~ \left ( \frac{c}{w_{j}} ~-~ \gamma ~c \right ) ~,
\nonumber \\
\vec{p}_{o}  &=& \left ( 1 ~-~	\frac{C'_{pr, 1}}{C'_{ext}}  \right ) ~
		 \frac{E_{i}}{c} ~\vec{e}_{1}  ~+~
		 \frac{C'_{pr, 1}}{C'_{ext}} ~
		 \frac{w_{1} ~E_{i}}{c^{2}} ~\gamma ~ \vec{v}
\nonumber \\
& &  -~  \sum_{j=2}^{3} ~ \frac{C'_{pr, j}}{C'_{ext}}  ~\frac{w_{1}
     ~E_{i}}{c^{2}} ~ \left ( c~ \frac{\vec{e}_{j}}{w_{j}}
     ~-~ \gamma ~ \vec{v}  \right )
\nonumber \\
& & -~ \frac{1}{c} \sum_{j=1}^{3} ~F'_{e, j} ~ \left ( c ~\frac{\vec{e}_{j}}{w_{j}}
    ~-~ \gamma	~ \vec{v} \right ) ~.
\end{eqnarray}

\subsection{Equation of motion}
In analogy with Eqs. (12), we have for the changes of energy and momentum of the
particle due to the interaction with electromagnetic radiation
\begin{eqnarray}\label{28}
\frac{d ~E}{d~ \tau} &=& E_{i}	~-~ E_{o} ~,
\nonumber \\
\frac{d ~\vec{p}}{d~ \tau} &=& \vec{p}_{i}  ~-~ \vec{p}_{o}  ~.
\end{eqnarray}
Putting Eqs. (27) into Eqs. (28), using also
$\vec{p}_{i}$ $=$ ( $E_{i}$ $/$ $c$ ) $\vec{e}_{1}$,
one easily obtains
\begin{eqnarray}\label{29}
\frac{d ~E / c}{d~ \tau} &=&  \sum_{j=1}^{3} \left (
      \frac{C'_{pr, j}}{C'_{ext}}  ~
      \frac{w_{1} ~ E_{i}}{c^{2}} ~+~ \frac{1}{c} ~F'_{e, j} \right )
      \left ( c ~ \frac{1}{w_{j}}  ~-~ \gamma ~c \right ) ~,
\nonumber \\
\frac{d ~\vec{p}}{d~ \tau} &=& \sum_{j=1}^{3} \left (
	\frac{C'_{pr, j}}{C'_{ext}} ~
	\frac{w_{1} ~ E_{i}}{c^{2}} ~+~ \frac{1}{c} ~F'_{e, j} \right )
       ~\left ( c ~\frac{\vec{e}_{j}}{w_{j}}
      ~-~ \gamma \vec{v}  \right )  ~.
\end{eqnarray}

Eq. (29) may be rewritten in terms of four-vectors:
\begin{eqnarray}\label{30}
\frac{d ~p^{\mu}}{d~ \tau} &=& \sum_{j=1}^{3} \left (
      \frac{C'_{pr, j}}{C'_{ext}}  ~
      \frac{w_{1} ~E_{i}}{c^{2}} ~+~ \frac{1}{c} ~F'_{e, j} \right )
      \left ( c ~ b_{j}^{\mu} ~-~ u^{\mu}  \right ) ~,
\end{eqnarray}
where $p^{\mu}$ is four-vector of the particle of mass $m$
\begin{equation}\label{31}
p^{\mu} = m~ u^{\mu} ~,
\end{equation}
four-vector of the world-velocity of the particle is
\begin{equation}\label{32}
u^{\mu} = ( \gamma ~c, \gamma ~ \vec{v} ) ~.
\end{equation}
We have also other four-vectors
\begin{equation}\label{33}
b_{j}^{\mu} = ( 1 / w_{j} , \vec{e}_{j} / w_{j} ) ~, ~~ j= 1, 2, 3 ~.
\end{equation}

It can be easily verified that:\\
i) the quantity $w~E_{i}$ is a scalar quantity
-- see first of Eqs. (19); \\
ii) Eq. (30) reduces to Eq. (10) for the case
of proper inertial frame of reference of the particle; \\
iii) Eq. (30) yields $d ~m / d~ \tau =$ 0.

We introduce
\begin{eqnarray}\label{34}
b_{j}^{0} &\equiv& 1 / w_{j} \approx 1 ~+~ \vec{v} \cdot \vec{e}'_{j} ~/~c  ~,
\nonumber \\
\vec{b_{j}} &\equiv& \vec{e}_{j} ~/~ w_{j} \approx \vec{e}'_{j} ~+~
						   \vec{v} ~/~c ~,
\nonumber \\
& &	  ~~j = 1,~2,~3
\end{eqnarray}
for the purpose of practical calculations. Physics of these relations
corresponds to aberration of light.
In general case, $b_{j}^{'~ \mu} = ( 1, \vec{e}'_{j} )$,
$b_{j}^{\mu} = \Lambda _{\nu}^{~~ \mu} b_{j}^{' ~\nu}$, $j \in \{
1, 2, 3 \}$, where $\Lambda _{\nu}^{~~ \mu}$ represents general
Lorentz transformation.
Eq. (34) is correct to the first order in $\vec{v}/c$, as it is presented.

We have derived an equation of motion for real dust particle under
the action of electromagnetic radiation (including thermal emission).
It is supposed that the
equation of motion is represented by Eqs. (11) and (12) in the
proper frame of reference of the particle. The final covariant
form is represented by Eq. (30), or using $E_{i} = w_{1} ~ S ~C'_{ext}$
(see Eqs. (11), (19) and (23)),
\begin{eqnarray}\label{35}
\frac{d ~p^{\mu}}{d~ \tau} &=& \sum_{j=1}^{3} \left (
      \frac{w_{1}^{2} ~S}{c^{2}} ~C'_{pr, j}  ~+~
      \frac{1}{c} ~F'_{e, j} \right )
      \left ( c ~ b_{j}^{\mu} ~-~ u^{\mu}  \right )  ~.
\end{eqnarray}

To first order in $\vec{v} / c$, Eqs. (34)-(35) yield
\begin{eqnarray}\label{36}
\frac{d~ \vec{v}}{d ~t} &=&  \frac{S}{m~c} ~ \sum_{j=1}^{3} ~C'_{pr, j}
      ~\left [	\left ( 1~-~ 2~
      \vec{v} \cdot \vec{e}_{1} / c ~+~
      \vec{v} \cdot \vec{e}_{j} / c \right ) ~ \vec{e}_{j}
      ~-~ \vec{v} / c \right ]	~+~
\nonumber   \\
& &   \frac{1}{m} ~ \sum_{j=1}^{3} ~F'_{e, j} ~\left [  \left ( 1~+~
      \frac{\vec{v} \cdot \vec{e}_{j}}{c} \right ) ~ \vec{e}_{j}
      ~-~ \frac{\vec{v}}{c} \right ]  ~,
\nonumber \\
\vec{e}_{j} &=& ( 1 ~-~ \vec{v} \cdot \vec{e}'_{j} / c ) ~ \vec{e}'_{j} ~+~
      \vec{v} / c ~~~, ~~j = 1, 2, 3~.
\end{eqnarray}
It is worth mentioning to stress that the values of radiation pressure
cross sections $C'_{pr, j}$, $j$ $=$ 1, 2, 3,
depend on particle's orientation with respect to
the incident radiation -- their values are time dependent.

It can be verified that Eq. (35) (or Eq. 36
within the accuracy to the first order in $\vec{v} / c$) yields as special
cases the situations discussed in Einstein (1905) and Robertson (1937).
However, Eqs. (11) and (27) are not consistent with the statement
presented by Poynting (1903), Robertson (1937), Wyatt and Whipple (1950)
and others about the process of reemission of perfectly absorbing spherical
dust particle (explanation is presented in Secs. 7.3.3 and 7.3.4).

\subsection{Continuous distribution of density flux of energy}
For a continuous frequency distribution of density flux of energy, we can write
\begin{eqnarray}\label{37}
\frac{d ~\vec{p'}}{d~ \tau} &=& \sum_{j=1}^{3} \left \{ \frac{1}{c} ~
     \int_{0}^{\infty} ~c~h ~\nu ' ~
     \frac{\partial n'}{\partial \nu '} ~ C'_{pr, j} ( \nu ') ~ d \nu ' ~+~
     F'_{e, j}  \right \} ~\vec{e}_{j} ' \equiv
\nonumber \\
&\equiv& \sum_{j=1}^{3} \left ( \frac{S'}{c} ~ \bar{C}'_{pr, j} ~+~
     F'_{e, j}  \right ) ~\vec{e'}_{j} ~.
\end{eqnarray}
Taking into account that concentration of photons fulfills
$n' = w_{1} ~n$ (Eq. (22)) and that Doppler effect yields
$\nu ' = w_{1}~ \nu$ (Eq. (17)), we have $\partial n' / \partial \nu '$ $=$
$\partial n / \partial \nu$. Lorentz transformation finally yields
\begin{eqnarray}\label{38}
\frac{d ~p^{\mu}}{d~ \tau} &=& \sum_{j=1}^{3} \left \{
     \frac{w_{1}^{2}}{c^{2}} ~ \int_{0}^{\infty} ~
     c~h ~ \frac{\partial n}{\partial \nu} ~\nu~ C'_{pr, j} ( w_{1} ~\nu ) ~
     d \nu ~+~ \frac{1}{c} ~F'_{e, j} \right \} \times
\nonumber \\
& &	\left ( c ~ b_{j}^{\mu} ~-~ u^{\mu}  \right )
\nonumber \\ &\equiv& \sum_{j=1}^{3} \left (
\frac{w_{1}^{2}~S}{c^{2}} ~ \bar{C}'_{pr, j} ~
     ~+~ \frac{1}{c} ~F'_{e, j} \right )
     \left ( c ~ b_{j}^{\mu} ~-~ u^{\mu}  \right ) ~.
\end{eqnarray}
As a consequence, $d m / d \tau =$ 0 (this corresponds to the condition
$d E' / d \tau =$ 0). As for the accuracy to the first order in $v/c$,
equation of motion of the type Eq. (36) holds if the substitution
$C'_{pr, j}$ $\rightarrow$ $\bar{C}'_{pr, j}$, $j$ $=$ 1, 2, 3 is done:
\begin{eqnarray}\label{39}
\frac{d~ \vec{v}}{d ~t} &=&  \frac{S}{m~c} ~ \sum_{j=1}^{3} ~\bar{C}'_{pr, j}
      ~\left [	\left ( 1~-~ 2~
      \vec{v} \cdot \vec{e}_{1} / c ~+~
      \vec{v} \cdot \vec{e}_{j} / c \right ) ~ \vec{e}_{j}
      ~-~ \vec{v} / c \right ]	~+~
\nonumber   \\
& &   \frac{1}{m} ~ \sum_{j=1}^{3} ~F'_{e, j} ~\left [  \left ( 1~+~
      \frac{\vec{v} \cdot \vec{e}_{j}}{c} \right ) ~ \vec{e}_{j}
      ~-~ \frac{\vec{v}}{c} \right ]  ~,
\nonumber \\
\vec{e}_{j} &=& ( 1 ~-~ \vec{v} \cdot \vec{e}'_{j} / c ) ~ \vec{e}'_{j} ~+~
      \vec{v} / c ~~~, ~~j = 1, 2, 3~.
\end{eqnarray}
The values of radiation pressure cross sections $\bar{C}'_{pr, j}$, $j$ $=$
1 to 3,  depend on particle's orientation with respect to
the incident radiation.

If the particle is not irradiated for a time interval, then
outgoing radiation due to the thermal emission is given by four-vector
( $E'_{o} / c$, $\vec{p}'_{o}$ ), where $E'_{o}$ is energy per unit time and
$\vec{p}'_{o}$ $=$ $-$ $\vec{F}'_{e}$ $\equiv$ $-$ $\sum_{j=1}^{3}$
$F'_{e, j}$ $\vec{e}'_{j}$ is momentum per unit time. Using transformations
represented by Eq. (14), or Eq. (26), and using also Eq. (24) together with
the fact that the four-force acting on the particle is
( $d ~p^{\mu}$ $/$ $d~ \tau$ ) $_{e}$ $=$ $-$ $p_{o}^{\mu}$, one obtains
\begin{equation}\label{40}
\left ( \frac{d ~p^{\mu}}{d~ \tau} \right ) _{e} =
  -~ \frac{E'_{o}}{c}~ \frac{u^{\mu}}{c} ~+~
  \sum_{j=1}^{3} ~F_{e, j}' \left ( b_{j}^{\mu} ~-~ \frac{u^{\mu}}{c} \right )~,
\end{equation}
instead of Eq. (38) and $E'_{o}$ is energy per unit time, which is lost
due to the thermal emission. Eq. (40) yields
\begin{eqnarray}\label{41}
\left ( \frac{d ~u^{\mu}}{d~ \tau} \right ) _{e} &=& \frac{1}{m} ~
  \sum_{j=1}^{3} ~F_{e, j}' \left ( b_{j}^{\mu} ~-~ \frac{u^{\mu}}{c} \right )~,
\nonumber \\
\left ( \frac{d m}{d \tau} \right ) _{e} &=& - ~\frac{E'_{o}}{c^{2}} ~.
\end{eqnarray}
Mass of the particle decreases due to the thermal emission, alone.

\section{Incoming and outgoing radiation -- covariant formulation}
We have treated the process of interaction between the electromagnetic
radiation and a dust grain in two ways: we have considered the
incoming/incident radiation and the outgoing radiation.
As we are interested in the relations
between the corresponding part of radiation and it's mass, we formulate
covariant forms for the two parts of radiation, in this section.

\subsection{Momentum-four vector for incoming radiation}
On the basis of Eq. (11) and Sec. 3.1, mainly Eqs. (18)-(23), and also
using the results of Sec. 3.4, we can write four-momentum of the
incoming (incident) radiation
\begin{equation}\label{42}
d p_{in}^{\mu} = \frac{w_{1}^{2} ~ S ~ \bar{C}'_{ext}}{c} ~
		 b_{1}^{\mu} ~ d \tau ~,
\end{equation}
where also Eqs. (25) (or 18) and (33) have to be used.

\subsection{Momentum-four vector for outgoing radiation}
On the basis of Eqs. (23), (25), (27), (32), (33) and the results of Sec. 3.4,
we can write for the energy and momentum of the outgoing radiation
\begin{eqnarray}\label{43}
\frac{d E_{out}}{c} &=& X~ \left \{ \frac{1}{w_{1}} ~-~ \sum_{j=1}^{3} \left (
       \frac{\bar{C}'_{pr, j}}{\bar{C}'_{ext}} ~+~ X^{-1} ~ F'_{e, j}
       \right ) ~ \left ( \frac{1}{w_{j}} - \gamma  \right ) \right \} ~
       d \tau ~,
\nonumber \\
d \vec{p}_{out} &=& X~ \left \{ \frac{\vec{e}_{1}}{w_{1}} ~-~ \sum_{j=1}^{3}
	      \left ( \frac{\bar{C}'_{pr, j}}{\bar{C}'_{ext}}
	      ~+~ X^{-1} ~ F'_{e, j} \right ) ~ \left (
	      \frac{\vec{e}_{j}}{w_{j}} - \gamma ~\frac{\vec{v}}{c} \right )
	      \right \} ~ d \tau ~,
\nonumber \\
X &\equiv& \frac{w_{1}^{2} ~ S~ \bar{C}'_{ext}}{c} ~.
\end{eqnarray}
Eq. (43) can be written in a short relativistically covariant form
\begin{eqnarray}\label{44}
d p_{out}^{\mu} &=& X~ \left \{ b_{1}^{\mu} ~-~ \sum_{j=1}^{3} \left (
	      \frac{\bar{C} '_{pr, j}}{\bar{C}' _{ext}} ~+~ X^{-1} ~ F'_{e, j}
	      \right ) ~ \left ( b_{j}^{\mu} - \frac{1}{c} ~u^{\mu} \right )
	      \right \} ~ d \tau ~,
\nonumber \\
X &\equiv& \frac{w_{1}^{2} ~ S ~ \bar{C}'_{ext}}{c} ~,
\nonumber \\
p_{out}^{\mu} &=& \left ( \frac{E_{out}}{c} , ~\vec{p}_{out} \right )
\end{eqnarray}
and Eqs. (25), (32) and (33) can be used.

\subsection{Relation between outgoing and incoming radiation}
Eq. (44) reads
\begin{eqnarray}\label{45}
d p_{out}^{\mu} &=& \frac{w_{1}^{2}~S~\bar{C}'_{ext}}{c} ~ \left \{
      b_{1}^{\mu} ~-~
      \sum_{j=1}^{3} \frac{\bar{C}'_{pr, j}}{\bar{C}'_{ext}}
      \left ( b_{j}^{\mu} ~-~ \frac{u^{\mu}}{c}  \right ) ~ \right \} ~d \tau
\nonumber \\
& & -~ \sum_{j=1}^{3}  F'_{e, j}  ~
      \left ( b_{j}^{\mu} ~-~ \frac{u^{\mu}}{c}  \right ) ~d \tau ~.
\end{eqnarray}
Eqs. (42) and (45) imediately yield
\begin{eqnarray}\label{46}
d p_{out}^{\mu} &=& \left (  1 ~-~
	\frac{\bar{C}'_{pr, 1}}{\bar{C}'_{ext}} \right ) ~
	d p_{in}^{\mu} ~+~
	\frac{w_{1}^{2}~S~\bar{C}'_{ext}}{c} ~
	\frac{\bar{C}'_{pr, 1}}{\bar{C}'_{ext}} ~
	\frac{u^{\mu}}{c} ~d \tau
\nonumber \\
& & -~ \frac{w_{1}^{2}~S~\bar{C}'_{ext}}{c} ~\sum_{j=2}^{3}
       \frac{\bar{C}'_{pr, j}}{\bar{C}'_{ext}} ~
       \left ( b_{j}^{\mu} ~-~ \frac{u^{\mu}}{c}  \right ) ~d \tau
\nonumber \\
& & -~ \sum_{j=1}^{3}  F'_{e, j}  ~
      \left ( b_{j}^{\mu} ~-~ \frac{u^{\mu}}{c}  \right ) ~d \tau ~.
\end{eqnarray}
Eq. (46) is covariant form of Eq. (11).

\subsection{Momentum-four vector for outgoing radiation -- thermal emission}
If the particle is not irradiated, then, in accordance with Eq. (40), we have
\begin{equation}\label{47}
\left [ \left ( d p_{out} \right ) _{e} \right ] ^{\mu}= \left \{
  \frac{E'_{o}}{c}~ \frac{u^{\mu}}{c} ~-~
  \sum_{j=1}^{3} ~F_{e, j}' \left ( b_{j}^{\mu} ~-~ \frac{u^{\mu}}{c} \right )
  \right \} ~ d \tau ~,
\end{equation}
for outgoing radiation due to the thermal emission.

\section{Radiation: Energy-mass relation}
Now, we are interested in the relations between the corresponding part
of radiation, incoming or outgoing, and it's mass.

\subsection{Incoming radiation}
Making an invariant of Eq. (42)
\begin{equation}\label{48}
\left ( c~d M_{in} \right ) ^{2} = d p_{in}^{\mu} ~d p_{in~\mu} ~,
\end{equation}
one immediately obtains, on the basis of Eqs. (25) and (33) --
$b_{1}^{\mu} ~b_{1~\mu}$ $=$ 0 --
\begin{equation}\label{49}
d M_{in} = 0 ~;
\end{equation}
a summation over repeated upper and lower indices is always implied, e. g.,
$d p_{in}^{\mu} ~d p_{in~\mu}$ $=$ $d p_{in}^{0} ~d p_{in~0}$ $+$
$d p_{in}^{1} ~d p_{in~1}$ $+$ $d p_{in}^{2} ~d p_{in~2}$ $+$
$d p_{in}^{3} ~d p_{in~3}$ $=$ $( d p_{in}^{0} )^{2}$ $-$
$( d p_{in}^{1} )^{2}$ $-$ $( d p_{in}^{2} )^{2}$ $-$ $( d p_{in}^{3} )^{2}$.

On the basis of Eqs. (42) and (49), we can summarize:
\begin{eqnarray}\label{50}
d E_{in} &=& \frac{w_{1}^{2}~S~\bar{C}'_{ext}}{c}  ~c~
	     \frac{1}{w_{1}} ~d \tau ~,
\nonumber \\
d \vec{p}_{in} &=& \frac{w_{1}^{2}~S~\bar{C}'_{ext}}{c}  ~
		   \frac{\vec{e}_{1}}{w_{1}} ~d \tau ~,
\nonumber \\
d M_{in} &=& 0 ~.
\end{eqnarray}
Eq. (49) states that energy of the incoming/incident radiation is nonzero,
while (the invariant) mass of the radiation is zero. The zeroness of the mass
of the incident radiation is understandable, since the radiation is produced by
the parallel flux of photons moving in the same direction and orientation
(see Okun 1989a, 1989b; Schr\"{o}der 1990, p. 108: "For the invariant
description of the inertial behaviour of a particle, only its rest mass
$m$ can be used, because it has the same value in all reference systems.").

\subsection{Outgoing radiation}
On the basis of Eqs. (25), (31)-(33) and (44), we can write
\begin{eqnarray}\label{51}
d p_{out}^{\mu} &=& \frac{w_{1}^{2}~S~\bar{C}'_{ext}}{c} ~
      \left \{b_{1}^{\mu} ~-~
      \sum_{j=1}^{3} \frac{\bar{C}'_{pr, j}}{\bar{C}'_{ext}}
      \left ( b_{j}^{\mu} ~-~ \frac{u^{\mu}}{c}  \right ) ~ \right \} ~d \tau
      ~-~
\nonumber \\
& & -~ \sum_{j=1}^{3}  F'_{e, j}  ~
      \left ( b_{j}^{\mu} ~-~ \frac{u^{\mu}}{c}  \right ) ~d \tau ~,
\nonumber \\
u_{\mu} ~b_{j}^{\mu} &=& c ~, ~~j = 1,~ 2,~ 3~,
\nonumber \\
u_{\mu} ~u^{\mu} &=& c^{2} ~,
\nonumber \\
b_{i ~\mu} ~b_{j}^{\mu} &=& 1 ~-~ \delta_{i j} ~,~~ i, j = 1~to~ 3~,
\end{eqnarray}
where $\delta_{i j}$ is Kronecker delta
($\delta_{i j}$ $=$ 1 if $i = j$, $\delta_{i j}$ $=$ 0 if $i \ne j$).

Making an invariant of Eq. (44) or of the first equation in Eqs. (51),
\begin{equation}\label{52}
\left ( c~d M_{out} \right ) ^{2} = d p_{out ~\mu} ~d p_{out}^{\mu} ~,
\end{equation}
one immediately obtains, on the basis of Eqs. (51) and (52),
\begin{eqnarray}\label{53}
d M_{out} &=& \frac{w_{1}^{2}~S~\bar{C}'_{ext}}{c^{2}} ~
	    \sqrt{ 2 \left (
	    \frac{\bar{C}'_{pr, 1}}{\bar{C}'_{ext}} ~+~ X^{-1} ~ F'_{e, 1}
	    \right ) ~-~ \sum_{j=1}^{3} \left (
	    \frac{\bar{C}'_{pr, j}}{\bar{C}'_{ext}} ~+~ X^{-1} ~ F'_{e, j}
	    \right )^{2} } ~ d \tau ~,
\nonumber \\
X &\equiv& \frac{w_{1}^{2} ~ S ~ \bar{C}'_{ext}}{c} ~.
\end{eqnarray}

On the basis of Eqs. (44) and (53), we can summarize:
\begin{eqnarray}\label{54}
d E_{out} &=& X~c~ \left \{ \frac{1}{w_{1}} ~-~ \sum_{j=1}^{3} \left (
	      \frac{\bar{C}'_{pr, j}}{\bar{C}'_{ext}} ~+~ X^{-1} ~ F'_{e, j}
	      \right ) ~ \left ( \frac{1}{w_{j}} - \gamma \right )
	      \right \} ~ d \tau ~,
\nonumber \\
d M_{out} &=& X ~\frac{1}{c} ~ \sqrt{ 2 \left (
	      \frac{\bar{C}'_{pr, 1}}{\bar{C}'_{ext}} ~+~
	      X^{-1} ~ F'_{e, 1} \right ) ~-~
	      \sum_{j=1}^{3} \left (
	      \frac{\bar{C}'_{pr, j}}{\bar{C}'_{ext}} ~+~
	      X^{-1} ~ F'_{e, j} \right )^{2} } ~ d \tau ~,
\nonumber \\
X &\equiv& \frac{w_{1}^{2} ~ S~ \bar{C}'_{ext}}{c} ~.
\end{eqnarray}
Eq. (54) states that energy of the outgoing radiation is nonzero,
and also (the invariant) mass of the radiation is nonzero. The procedure
of the calculation of the mass is consistent with the procedure discussed
by Okun (1989a, 1989b).

Eqs. (51) and (54) yield
\begin{eqnarray}\label{55}
d E_{out} &=& \gamma_{out} ~\left ( d M_{out} \right ) ~ c^{2} ~,
\nonumber \\
d \vec{p}_{out} &=& \gamma_{out} ~
	       \left ( d M_{out} \right ) ~\vec{v}_{out} ~,
\nonumber \\
\vec{v}_{out} &=& \left \{ \frac{\vec{e}_{1}}{w_{1}} ~-~ \sum_{j=1}^{3} \left (
	      \frac{\bar{C}'_{pr, j}}{\bar{C}'_{ext}} ~+~ X^{-1} ~ F'_{e, j}
	      \right ) ~ \left ( \frac{\vec{e}_{j}}{w_{j}} - \gamma
	      \frac{\vec{v}}{c} \right )  \right \} ~
	      \left ( ZC \right ) ^{-1} ~ c ~,
\nonumber \\
\gamma_{out} &=& \frac{ZC}{ZM} ~,
\nonumber \\
ZC &=& \frac{1}{w_{1}} ~-~ \sum_{j=1}^{3} \left (
	      \frac{\bar{C}'_{pr, j}}{\bar{C}'_{ext}} ~+~ X^{-1} ~ F'_{e, j}
	      \right ) ~ \left ( \frac{1}{w_{j}} - \gamma \right ) ~,
\nonumber \\
ZM &=&	\sqrt{ 2 \left ( \frac{\bar{C}'_{pr, 1}}{\bar{C}'_{ext}} ~+~
	      X^{-1} ~ F'_{e, 1} \right ) ~-~
	      \sum_{j=1}^{3} \left (
	      \frac{\bar{C}'_{pr, j}}{\bar{C}'_{ext}} ~+~
	      X^{-1} ~ F'_{e, j} \right )^{2} } ~.
\nonumber \\
X &\equiv& \frac{w_{1}^{2} ~ S~ \bar{C}'_{ext}}{c} ~.
\end{eqnarray}
We have obtained the standard formulae: energy equals $\gamma$ $\times$ mass
$\times$ $c^{2}$, momentum equals $\gamma$ $\times$ mass $\times$ velocity.

\subsection{Outgoing radiation -- thermal emission}
Making an invariant of Eq. (47)
\begin{equation}\label{56}
\left [ \left ( c~d M_{out} \right )_{e} \right ] ^{2} = \left [
    \left ( d p_{out} \right )_{e} \right ] _{\mu} ~
    \left [ \left ( d p_{out} \right )_{e} \right ] ^{\mu} ~,
\end{equation}
one easily obtains
\begin{equation}\label{57}
\left ( d M_{out} \right )_{e} = \sqrt{ \left ( \frac{E'_{o}}{c^{2}} \right )^{2}
	 ~-~ \sum_{j=1}^{3} \left ( \frac{F'_{e, j}}{c} \right )^{2}} ~ d \tau ~.
\end{equation}

Eqs. (47) and (57) yield
\begin{eqnarray}\label{58}
\left ( d E_{out} \right )_{e} &=& \left \{ \gamma ~E'_{o} ~-~
	 \sum_{j=1}^{3} c~ F'_{e, j} \left ( \frac{1}{w_{j}} ~-~ \gamma
	 \right ) \right \} ~ d \tau ~,
\nonumber \\
\left ( d M_{out} \right )_{e} &=& \sqrt{ \left ( \frac{E'_{o}}{c^{2}} \right )^{2}
	 ~-~ \sum_{j=1}^{3} \left ( \frac{F'_{e, j}}{c} \right )^{2}} ~ d \tau ~,
\nonumber \\
w_{j} &=& \gamma ~ \left ( 1 ~-~ \frac{\vec{v} \cdot \vec{e}_{j}}{c} \right ) ~,~~
	  j = 1, 2, 3 ~,
\end{eqnarray}
where also Eq. (25) was used.

Eqs. (47) and (58) yield
\begin{eqnarray}\label{59}
\left ( d E_{out} \right )_{e} &=& \gamma_{e} ~
	       \left ( d M_{out} \right )_{e} ~c^{2} ~,
\nonumber \\
\left ( d \vec{p}_{out} \right )_{e} &=& \gamma_{e} ~
	       \left ( d M_{out} \right )_{e} ~\vec{v}_{e} ~,
\nonumber \\
\vec{v}_{e} &=& \left \{ \frac{E'_{o}}{c} ~ \gamma ~ \frac{\vec{v}}{c} ~-~
	 \sum_{j=1}^{3} c~ F'_{e, j} \left ( \frac{\vec{e}_{j}}{w_{j}} ~-~
	 \gamma ~\frac{\vec{v}}{c} \right ) \right \} ~
	 \left ( ZC \right )_{e} ^{-1} ~ c^{2} ~,
\nonumber \\
\gamma_{e} &=& \frac{\left ( ZC \right )_{e}}{\left ( ZM \right )_{e}} ~,
\nonumber \\
\left ( ZC \right )_{e} &=& \gamma ~E'_{o} ~-~
	 \sum_{j=1}^{3} c~ F'_{e, j} \left ( \frac{1}{w_{j}} ~-~ \gamma
	 \right ) ~,
\nonumber \\
\left ( ZM \right )_{e} &=&
	     \sqrt{ \left ( E'_{o} \right )^{2}
	 ~-~ \sum_{j=1}^{3} \left ( c~ F'_{e, j} \right )^{2} }  ~.
\end{eqnarray}
We have obtained standard formulae: energy equals $\gamma$ $\times$ mass
$\times$ $c^{2}$, momentum equals $\gamma$ $\times$ mass $\times$ velocity.

\section{Incoming and outgoing radiation -- change of particle's mass}
As it was already pointed out just below Eq. (38), mass of the particle in the
whole process of interaction with the radiation is conserved, i.e. $d m / d \tau$
$=$ 0; if the particle is not irradiated, then Eqs. (40)-(41) hold.
Now, we are interested in the effect of incoming and outgoing radiation
on the mass of the particle.

\subsection{Incoming radiation}
On the basis of Eq. (42), we can write for the change of four-momentum of the
particle due to the incoming radiation
\begin{eqnarray}\label{60}
\frac{d p^{\mu}}{d \tau} &=& \frac{d p_{in}^{\mu}}{d \tau} ~,
\nonumber \\
\frac{d p_{in}^{\mu}}{d \tau} &=&
	  \frac{w_{1}^{2} ~ S ~ \bar{C}'_{ext}}{c} ~b_{1}^{\mu} ~,
\end{eqnarray}
which leads to the result
\begin{equation}\label{61}
\left ( \frac{d m}{d \tau} \right ) _{in} =
		     \frac{w_{1}^{2} ~ S ~ \bar{C}'_{ext}}{c} ~
		     \frac{1}{c} ~.
\end{equation}
Comparison of Eq. (61) with Eq. (50) yields
\begin{equation}\label{62}
\frac{d E_{in}}{d \tau} =
\left ( \frac{d m}{d \tau} \right ) _{in} ~c^{2} ~ \frac{1}{w_{1}}
\end{equation}
and
\begin{equation}\label{63}
\frac{d \vec{p}_{in}}{d \tau} = \left ( \frac{d m}{d \tau} \right )_{in} ~
				c ~\frac{\vec{e}_{1}}{w_{1}} ~.
\end{equation}

\subsection{Outgoing radiation}
On the basis of Eq. (51), we can write for the change of four-momentum of the
particle due to the outgoing radiation
\begin{eqnarray}\label{64}
\frac{d p^{\mu}}{d \tau} &=& -~ \frac{d p_{out}^{\mu}}{d \tau} ~,
\nonumber \\
\frac{d p_{out}^{\mu}}{d \tau} &=&
		 \frac{w_{1}^{2}~S~\bar{C}'_{ext}}{c} ~ \left \{
		 b_{1}^{\mu} ~-~
      \sum_{j=1}^{3} \frac{\bar{C}'_{pr, j}}{\bar{C}'_{ext}}
      \left ( b_{j}^{\mu} ~-~ \frac{u^{\mu}}{c}  \right ) ~ \right \}
\nonumber \\
& & -~ \sum_{j=1}^{3}  F'_{e, j}  ~
      \left ( b_{j}^{\mu} ~-~ \frac{u^{\mu}}{c}  \right )  ~.
\end{eqnarray}
This leads to the result
\begin{equation}\label{65}
\left ( \frac{d m}{d \tau} \right ) _{out} = -~
		     \frac{w_{1}^{2} ~ S ~ \bar{C}'_{ext}}{c} ~
		     \frac{1}{c} ~.
\end{equation}
According to Eqs. (61) and (65):
$( d m / d \tau ) _{out}$ $=$ $-$ $( d m / d \tau ) _{in}$.
Comparison of Eq. (65) with Eq. (54) yields
\begin{eqnarray}\label{66}
\frac{d E_{out}}{d \tau} &=& -~\left \{ \frac{1}{w_{1}} ~-~ \sum_{j=1}^{3}
      \left ( \frac{\bar{C}'_{pr, j}}{\bar{C}'_{ext}} ~+~ X^{-1} ~ F'_{e, j}
      \right ) ~ \left ( \frac{1}{w_{j}} - \gamma \right )
      \right \} ~\left ( \frac{d m}{d \tau} \right ) _{out} ~c^{2} ~,
\nonumber \\
\frac{d \vec{p}_{out}}{d \tau} &=& -~\left \{
	      \frac{\vec{e}_{1}}{w_{1}} ~-~ \sum_{j=1}^{3} \left (
	      \frac{\bar{C}'_{pr, j}}{\bar{C}'_{ext}} ~+~ X^{-1} ~ F'_{e, j}
	      \right ) ~ \left ( \frac{\vec{e}_{1}}{w_{j}} - \gamma ~
	      \frac{\vec{v}}{c} \right )
	      \right \} ~\left ( \frac{d m}{d \tau} \right ) _{out} ~c ~,
\nonumber \\
X &\equiv& \frac{w_{1}^{2} ~ S~ \bar{C}'_{ext}}{c} ~.
\end{eqnarray}

\subsection{Thermal emission}
As for the thermal emission alone, the relevant result is given by Eqs.
(40)-(41):
\begin{equation}\label{67}
\left ( \frac{d m}{d \tau} \right  )_{e} = - \frac{E'_{o}}{c^{2}} ~.
\end{equation}
Eqs. (40)-(41) immediately yield for the energy and momentum of the
outgoing radiation:
\begin{eqnarray}\label{68}
\frac{d E_{o}}{d \tau} &=& -~\gamma ~\left ( \frac{d m}{d \tau} \right ) _{e}~
	      c^{2}  ~-~ c~ \sum_{j=1}^{3} F'_{e, j} ~\left (
	      \frac{1}{w_{j}} - \gamma \right ) ~,
\nonumber \\
\frac{d \vec{p}_{o}}{d \tau} &=& -~\gamma ~
	      \left ( \frac{d m}{d \tau} \right ) _{e}~ \vec{v} ~-~
	      \sum_{j=1}^{3} F'_{e, j} ~\left (
	      \frac{\vec{e}_{1}}{w_{j}} - \gamma ~
	      \frac{\vec{v}}{c} \right ) ~.
\end{eqnarray}
Eq. (68) yields the standard formulae: energy equals $\gamma$ $\times$ mass
$\times$ $c^{2}$, momentum equals $\gamma$ $\times$ mass $\times$ velocity
in the case $F'_{e, j}$ $=$ 0, $j$ $=$ 1 to 3.

\section{Discussion}

\subsection{Physics of $p_{in}^{\mu}$}
Eq. (11) is physically different from Eq. (11) in Kla\v{c}ka (2004) and
also from Eq. (122) in Kla\v{c}ka (1992). In order to understand the
physics behind the formulae, we will consider a simple case of
spherically symmetric particle fulfilling $\bar{C}'_{pr, 2}$ $=$
$\bar{C}'_{pr, 3}$ $=$ $\bar{C}'_{abs}$ $=$ $F'_{e, 1}$ $=$
$F'_{e, 2}$ $=$ $F'_{e, 3}$ $=$ 0, now.

For the sake of brevity, we will use dimensionless efficiency factors $Q'_{x}$
instead of cross sections $C'_{x}$: $C'_{x} = Q'_{x} ~ A'$, where $A'$
is geometrical cross section of the spherical particle, and,
we will not use primed quantities.

Let a beam of homogeneous incoming photons, interacting with the spherical
particle, consists of $N$ photons each of which carries energy $\varepsilon$.
The total incoming momentum is $N$ ($\varepsilon / c$) $\vec{e}$ and the
elastically scattered photons are characterized by the total
outgoing momentum ($\varepsilon / c$) $\sum_{j=1}^{N}$ $\vec{e}_{j}$;
momentum of an individual photon is ($\varepsilon / c$) $\vec{e}_{j}$,
$j$ $=$ 1 to $N$. The momentum of the particle is given by conservation
of the total momentum:
$N$ ($\varepsilon / c$) $Q_{pr, 1}$ $\vec{e}$ $=$
$N$ ($\varepsilon / c$) $\vec{e}$ $-$
($\varepsilon / c$) $\sum_{j=1}^{N}$ $\vec{e}_{j}$, according to
Kla\v{c}ka (1992, 2004). This yields
$Q_{pr, 1}$ $\vec{e}$ $=$ $\vec{e}$ $-$
$\sum_{j=1}^{N}$ $\vec{e}_{j}$ $/$ $N$. Taking into account that the last term
is an average which can be written as $< \cos \Theta >$ $\vec{e}$, we finally
obtain $Q_{pr, 1}$ $=$ 1 $-$ $< \cos \Theta >$. However, this is not correct.
We know that $Q_{pr, 1}$ $A$ $\equiv$ $C_{pr, 1}$ $=$ $C_{ext}$ $-$
$< \cos \Theta >$ $C_{sca}$
and $Q_{ext}$ $A$ $=$ $Q_{abs}$ $A$ $+$ $Q_{sca}$ $A$ $=$ $Q_{sca}$ $A$,
since we deal with the case $Q_{abs}$ $=$ 0, in our example. Thus,
$Q_{pr, 1}$ $=$  $Q_{ext}$ ( 1 $-$ $< \cos \Theta >$ ) and the correct form of
the conservation of the total momentum is
$N$ ($\varepsilon / c$) $Q_{pr, 1}$ $\vec{e}$ $=$ $Q_{ext}$ [
$N$ ($\varepsilon / c$) $\vec{e}$ $-$ ($\varepsilon / c$)
$\sum_{j=1}^{N}$ $\vec{e}_{j}$ ] .... ($*$). This explains why the ratios of the
pressure terms to the extinction term are present in Eq. (11), and, also,
why the extinction cross section is present in energy in Eq. (11). The
equation ($*$) is fully consistent with Eq. (11): $\vec{p}'_{i}$ $=$
( $C'_{ext} / A'$ ) $\dot{N}'$ ( $\varepsilon ' / c$ )
$\vec{e}'$, $\vec{p}'_{o}$ $=$
[1 $-$ ( $C'_{pr, 1}$ $/$ $C'_{ext}$ ) ] $\vec{p}'_{i}$,
where the dot denotes differentiation with respect to time. All is consistent
with Eq. (5), too.

\subsection{Center-of-momentum}
Short comment on Eqs. (55) and (59). We have defined velocities $\vec{v}_{out}$
and $\vec{v}_{e}$ for radiation in Eqs. (55) and (59). This in reality
corresponds to the fact that we can introduce
{\it center-of-momentum frame}, where the total momentum vanishes.
Let $\vec{v}_{c}$ be the velocity of the center-of-momentum frame
with respect to the reference frame in which an object
(radiation, particle) of a mass $dM$ moves with velocity $\vec{v}$. Then
\begin{eqnarray}\label{69}
d E &=& \gamma_{c} ~\left ( d M \right ) ~ c^{2} ~,
\nonumber \\
d \vec{p} &=& \gamma_{c} ~\left ( d M \right ) ~ \vec{v}_{c} ~,
\nonumber \\
\vec{v}_{c} &=& c^{2} ~ \frac{d \vec{p}}{d E}
\end{eqnarray}
and {\it center-of-inertia} of the system of the outgoing photons, defined
as the point with radius vector $\vec{R}_{S} (t)$ $=$
( $\sum_{i} E_{i} (t) \vec{r}_{i} (t)$ ) $/$ ( $\sum_{i} E_{i} (t)$ ),
moves with the velocity $\vec{v}_{c}$ given by Eq. (69) (see, e. g.,
Landau and Lifshitz 2005: $\S$14, pp. 42-45, or, Ferraro 2007:
$\S$6.5, pp. 146-148, and $\S$6.7, pp. 153-155).

\subsection{Special case}
Now, we will deal with the special case
$\bar{C}'_{pr, 2}$ $=$ $\bar{C}'_{pr, 3}$ $=$
$F'_{e, j}$ $=$ 0 for $j$ $=$ 1, 2, 3. This case corresponds to secular
decrease of semi-major axis and eccentricity of neutral dust grain in the
gravitational and electromagnetic fields of a central star, e. g., Sun,
if the cross section $\bar{C}'_{pr, 1}$ is not a function of time
(see Secs. 6.1 and 6.2 in Kla\v{c}ka 2004).
Thus, we can call this effect as the (generalized) Poynting-Robertson effect.
As in Sec. 6.1, we will use dimensionless efficiency factors $\bar{Q}'_{pr, 1}$,
$\bar{Q}'_{ext}$ instead of cross sections $\bar{C}'_{pr, 1}$, $\bar{C}'_{ext}$:
$\bar{Q}'_{pr, 1}$ $=$ $\bar{C}'_{pr, 1}$ $/$ $A'$,
$\bar{Q}'_{ext}$ $=$ $\bar{C}'_{ext}$ $/$ $A'$, where $A'$ is geometrical
cross section of a sphere of volume equal to the volume of the particle.
In reality, the case treated in this section holds only for particles
with spherically symmetric distribution of mass. Dimensionless efficiency
factors can be calculated on the basis of Mie theory (Mie 1908, Bohren and
Huffman 1983).

Equation of motion for the discussed case is given by reduction of Eq. (38):
\begin{eqnarray}\label{70}
\frac{d ~p^{\mu}}{d~ \tau} &=&
		\frac{w_{1}^{2}~S~A'}{c^{2}} ~ \bar{Q}'_{pr, 1} ~
		\left ( c ~ b_{1}^{\mu} ~-~ u^{\mu}  \right ) ~,
\nonumber \\
b_{1}^{\mu} &=& ( 1 / w_{1} , \vec{e}_{1} / w_{1} ) ~,
\nonumber \\
w_{1} &=& \gamma ~ ( 1 ~-~ \vec{v} \cdot \vec{e}_{1} / c ) ~,
\end{eqnarray}
where also Eqs. (18) and (33) were used.
As a consequence, $d m / d \tau =$ 0 (this corresponds to the condition
$d E' / d \tau =$ 0, where $E'$ is energy of the particle measured in the
rest frame of the particle).
If the particle is not irradiated, then only thermal emission may exist
and one has to use
\begin{equation}\label{71}
\left ( \frac{d ~u^{\mu}}{d~ \tau} \right ) _{e} = 0~,
\end{equation}
instead of Eq. (70) (see Eq. 41).

To first order in $\vec{v} / c$, Eq. (70) yields
\begin{eqnarray}\label{72}
\frac{d~ \vec{v}}{d ~t} &=&  \frac{S~A'}{m~c} ~ \bar{Q}'_{pr, 1} ~ \left \{
      \left ( 1~-~ \frac{\vec{v} \cdot \vec{e}_{1}}{c}	\right ) ~ \vec{e}_{1}
      ~-~ \frac{\vec{v}}{c} \right \}
\nonumber \\
\vec{e}_{1} &=& \left ( 1 ~-~ \frac{\vec{v} \cdot \vec{e}'_{1}}{c} \right )
      ~ \vec{e}'_{1} ~+~ \frac{\vec{v}}{c} ~.
\end{eqnarray}
It is worth mentioning to stress that the value $\bar{Q}'_{pr, 1}$
does not depend on particle's orientation with respect to
the incident radiation, but it's value may be time dependent, e. g.,
when optical properties change with the particle's distance from the
central star (Kla\v{c}ka {\it et al.} 2007).

The total process of interaction can be written in the form of the
following equations (energies and momenta per unit time):
\begin{eqnarray}\label{73}
E_{o} ' &=& E_{i} ' = S'~A'~\bar{Q}'_{ext}  ~,
\nonumber \\
\vec{p}_{o} ' &=& ( 1 ~-~ \frac{\bar{Q}'_{pr, 1}}{\bar{Q}'_{ext}} ) ~
		  \vec{p}_{i} ' ~,
\nonumber \\
\vec{p}_{i} ' &=& ( E_{i} ' / c ) ~ \vec{e}_{1} ' ~,
\end{eqnarray}
where the index $''i''$ represents the incoming (incident) radiation,
beam of photons, the index $''o''$ represents the outgoing radiation.
Eq. (73) is a fundamental condition for the validity of the
Poynting-Robertson effect and it is a new condition.
It differs from the condition $\vec{p}_{o} '$ $=$
( 1 $-$ $\bar{Q}'_{pr, 1}$ ) $\vec{p}_{i} '$ consistent with
the statements presented in Poynting (1903), Robertson (1937),
Wyatt and Whipple (1950), Burns {\it et al.} (1979; Eqs. 10 and 11)
and in other papers. The condition represented by Eq. (73) does not exist
in the literature and any statement about the outgoing radiation,
published up to now, is not consistent with Eq. (73).

Covariant form of Eq. (73) can be obtained from Eq. (46):
\begin{eqnarray}\label{74}
\frac{d p_{out}^{\mu}}{d \tau} &=& \left (  1 ~-~ \frac{\bar{Q}'_{pr, 1}}{\bar{Q}'_{ext}} \right )
	~ \frac{d p_{in}^{\mu}}{d \tau} ~+~
	\frac{w_{1}^{2}~S~A'~\bar{Q}'_{ext}}{c} ~
	\frac{\bar{Q}'_{pr, 1}}{\bar{Q}'_{ext}} ~
	\frac{u^{\mu}}{c}
\end{eqnarray}
and incoming four-momentum of the radiation is given by Eq. (42):
\begin{equation}\label{75}
d p_{in}^{\mu} = \frac{w_{1}^{2} ~ S ~A'~ \bar{Q}'_{ext}}{c} ~b_{1}^{\mu} ~
		  d \tau ~.
\end{equation}

Momentum-four vector for the outgoing radiation
can be written in a short relativistically covariant form, on the basis
of Eq. (44), as follows:
\begin{eqnarray}\label{76}
\frac{d p_{out}^{\mu}}{d \tau} &=& X~ \left \{ b_{1}^{\mu} ~-~
	      \frac{\bar{Q} '_{pr, 1}}{\bar{Q}' _{ext}}
	     ~ \left ( b_{1}^{\mu} - \frac{u^{\mu}}{c} \right )
	      \right \}  ~,
\nonumber \\
X &\equiv& \frac{w_{1}^{2} ~ S ~A' ~\bar{Q}'_{ext}}{c} ~,
\nonumber \\
p_{out}^{\mu} &=& \left ( \frac{E_{out}}{c} , ~\vec{p}_{out} \right ) ~.
\end{eqnarray}
and Eqs. (18), (32) and (33) can be used.

\subsubsection{Energy-mass relation for radiation}
As for the incoming radiation, the relevant statements are represented by
Eqs. (42) and (50), as for the four-momentum, and also, as for the energy and
the mass of the radiation.

Now, we will treat the outgoing radiation. Eq. (51) reduces to (see also Eq. 76)
\begin{eqnarray}\label{77}
d p_{out}^{\mu} &=& \frac{w_{1}^{2}~S~A'~\bar{Q}'_{ext}}{c} ~ \left \{
      \left ( 1 ~-~ \frac{\bar{Q}'_{pr, 1}}{\bar{Q}'_{ext}}  \right ) ~
      b_{1}^{\mu} ~+~ \frac{\bar{Q}'_{pr, 1}}{\bar{Q}'_{ext}}  ~
      \frac{u^{\mu}}{c}  ~ \right \} ~d \tau ~,
\nonumber \\
u_{\mu} ~b_{1}^{\mu} &=& c ~,
\nonumber \\
u_{\mu} ~u^{\mu} &=& c^{2} ~,
\nonumber \\
b_{1 ~\mu} ~b_{1}^{\mu} &=& 0~.
\end{eqnarray}
Definition of the invariant mass represented by Eq. (52),
$d M_{out}$ $=$ $\sqrt{d p_{out ~\mu} ~d p_{out}^{\mu}}$ $/$ $c^{2}$,
leads to
\begin{eqnarray}\label{78}
d M_{out} &=& \frac{w_{1}^{2}~S~A'~\bar{Q}'_{ext}}{c^{2}} ~
	    \sqrt{ \frac{\bar{Q}'_{pr, 1}}{\bar{Q}'_{ext}} ~ \left ( 2 ~-~
	    \frac{\bar{Q}'_{pr, 1}}{\bar{Q}'_{ext}} \right )} ~ d \tau ~,
\end{eqnarray}
if one uses results presented in Eq. (77); compare with Eq. (53).
As a consequence of Eq. (78), one obtains a simple
statement:
\begin{equation}\label{79}
0 < \frac{\bar{Q}'_{pr, 1}}{\bar{Q}'_{ext}} \le 2 ~.
\end{equation}

Eqs. (77) and (78) yield
\begin{eqnarray}\label{80}
d E_{out} &=& \frac{\left ( ZC \right )_{1}}{\left ( ZM \right )_{1}} ~
	      \left ( d M_{out} \right ) ~ c^{2} ~,
\nonumber \\
d \vec{p}_{out} &=& \frac{1}{\left ( ZM \right )_{1}} ~ \left \{
      \left ( 1 ~-~ \frac{\bar{Q}'_{pr, 1}}{\bar{Q}'_{ext}}  \right ) ~
      \frac{\vec{e}_{1}}{w_{1}} ~+~
      \frac{\bar{Q}'_{pr, 1}}{\bar{Q}'_{ext}} ~\gamma ~\frac{\vec{v}}{c}
      \right \} ~ \left ( d M_{out} \right ) ~ c ~,
\nonumber \\
\left ( ZC \right )_{1} &=&
      \left ( 1 ~-~ \frac{\bar{Q}'_{pr, 1}}{\bar{Q}'_{ext}}  \right ) ~
      \frac{1}{w_{1}} ~+~
	      \frac{\bar{Q}'_{pr, 1}}{\bar{Q}'_{ext}} ~\gamma ~,
\nonumber \\
\left ( ZM \right )_{1} &=&
	    \sqrt{ \frac{\bar{Q}'_{pr, 1}}{\bar{Q}'_{ext}} ~ \left ( 2 ~-~
	    \frac{\bar{Q}'_{pr, 1}}{\bar{Q}'_{ext}} \right )} ~,
\nonumber \\
w_{1} &=& \gamma ~ \left ( 1 ~-~ \frac{\vec{v} \cdot \vec{e}_{1}}{c} \right ) ~,
\end{eqnarray}
see also Eq. (55).

As for the center-of-momentum frame, Eqs. (69) and (80) lead to
\begin{eqnarray}\label{81}
\vec{v}_{c} &=& \frac{\bar{Q}' ~\gamma~\vec{v} ~+~
		\left ( 1 ~-~ \bar{Q}' \right ) ~c ~\vec{e}_{1} / w_{1} }{
		\bar{Q}' ~\gamma ~+~ \left ( 1 ~-~ \bar{Q}' \right ) / w_{1} } ~,
\nonumber \\
\bar{Q}' &\equiv& \frac{\bar{Q}'_{pr, 1}}{\bar{Q}'_{ext}} ~.
\end{eqnarray}
It can be verified that Eq. (81) yields
\begin{eqnarray}\label{82}
\vec{v}_{c} &=& \left (  1 ~-~ \bar{Q}' \right ) ~ c ~\vec{e}_{1} ~, ~~~
		if ~~\vec{v} = 0~,
\nonumber \\
\vec{v} &=& -~ \left ( 1~-~ \bar{Q}' \right ) ~ c ~\vec{e}_{1} ~, ~~~
		if ~~\vec{v}_{c} = 0~,
\nonumber \\
\vec{v}_{c} &=& \vec{v} ~, ~~~if ~~\bar{Q}' = 1~,
\nonumber \\
\bar{Q}' &\equiv& \frac{\bar{Q}'_{pr, 1}}{\bar{Q}'_{ext}} ~.
\end{eqnarray}
The case $\vec{v}$ $=$ $-$ ( 1 $-$ $\bar{Q}'_{pr, 1}$ $/$ $\bar{Q}'_{ext}$ )
$c$ $\vec{e}_{1}$ yields $dE_{out}$ $=$ ( $dM_{out}$ ) $c^{2}$, as can be
verified on the basis of Eq. (80).

If the particle is not irradiated, then the general Eqs. (58)-(59) reduce to
\begin{eqnarray}\label{83}
\left ( d E_{out} \right )_{e} &=&
		  \gamma ~\left ( d M_{out} \right )_{e} ~ c^{2} ~,
\nonumber \\
\left ( d \vec{p}_{out} \right )_{e} &=&
		  \gamma ~\left ( d M_{out} \right )_{e} ~ \vec{v} ~,
\nonumber \\
\left ( d M_{out} \right )_{e} &=& \frac{E'_{o}}{c^{2}} ~d \tau ~.
\end{eqnarray}
Moreover, Eq. (59) or comparison of Eq. (83) with Eq. (69) immediately gives
\begin{equation}\label{84}
\vec{v}_{c} = \vec{v} ~.
\end{equation}
Eq. (84) states that the center-of-momentum frame for thermally emitted
radiation is in the center of the spherical particle.

\subsubsection{Energy-mass relation for the particle}
At first, we will treat the incoming radiation.
On the basis of Eqs. (61)-(62), we have
\begin{eqnarray}\label{85}
\frac{d E_{in}}{d \tau} &=& \frac{1}{w_{1}} ~ \left (
			  \frac{d m}{d \tau} \right )_{in} ~ c^{2} ~,
\nonumber \\
\left ( \frac{d m}{d \tau} \right ) _{in} &=&
		     \frac{w_{1}^{2} ~ S ~A'~ \bar{Q}'_{ext}}{c^{2}} ~.
\end{eqnarray}
This is not equivalent to the change of mass of the particle
according to the "definition" $c^{-2}$ $d E_{in}$ $/$ $d \tau$
("mass equals energy divided by $c^{2}$"),
mainly if one takes into account that
Eqs. (60)-(61) and (69) lead to $\vec{v}_{c}$ $=$ $c$ $\vec{e}_{1}$:
the relation between the momentum of the incoming radiation and the particle's
mass is (see Eq. 63)
\begin{eqnarray}\label{86}
\frac{d \vec{p}_{in}}{d \tau} &=& \frac{1}{w_{1}} ~
				\left ( \frac{d m}{d \tau} \right )_{in} ~
				c ~ \vec{e}_{1} ~,
\nonumber \\
\left ( \frac{d m}{d \tau} \right ) _{in} &=&
		     \frac{w_{1}^{2} ~ S ~A'~ \bar{Q}'_{ext}}{c^{2}} ~.
\end{eqnarray}
We also remind of Eq. (49) for the mass of the incoming radiation:
$dM_{in}$ $/$ $d \tau$ $=$ 0.

For the outgoing radiation, Eq. (66) reduces, for the spherically
distributed mass, to
\begin{eqnarray}\label{87}
\frac{d E_{out}}{d \tau} &=& - ~ \left \{ \left ( 1 ~-~
			  \frac{\bar{Q}'_{pr, 1}}{\bar{Q}'_{ext}} \right ) ~
			  \frac{1}{w_{1}} ~+~
		  \frac{\bar{Q}'_{pr, 1}}{\bar{Q}'_{ext}} ~\gamma \right \} ~
		  \left ( \frac{d m}{d \tau} \right ) _{out} ~c^{2} ~,
\nonumber \\
\left ( \frac{d m}{d \tau} \right ) _{out} &=& -~
		     \frac{w_{1}^{2} ~ S ~A'~ \bar{Q}'_{ext}}{c^{2}} ~.
\end{eqnarray}
This is not equivalent to the change of mass of the particle
according to the "definition"
$\gamma^{-1} c^{-2} d E_{out} / d \tau$. Changes of the particle's mass
and outgoing momentum are related as follows:
\begin{eqnarray}\label{88}
\frac{d \vec{p}_{out}}{d \tau} &=& -~ \left ( \frac{d m}{d \tau} \right )_{out}
			   ~\left \{ \left ( 1 ~-~
			   \frac{\bar{Q}'_{pr, 1}}{\bar{Q}'_{ext}} \right ) ~
			   \frac{\vec{e}_{1}}{w_{1}} ~+~
			   \frac{\bar{Q}'_{pr, 1}}{\bar{Q}'_{ext}} ~\gamma ~
			   \frac{\vec{v}}{c} \right \} ~c ~,
\nonumber \\
\left ( \frac{d m}{d \tau} \right ) _{out} &=& -~
		     \frac{w_{1}^{2} ~ S ~A'~ \bar{Q}'_{ext}}{c^{2}} ~.
\end{eqnarray}
Eqs. (87)-(88) yield, together with equation analogous to Eq. (69), that
$\vec{v}_{c}$ is given by Eq. (81), and, thus, we can write
\begin{eqnarray}\label{89}
\frac{d E_{out}}{d \tau} &=& -~ \gamma_{c} ~
		  \sqrt{\bar{Q}' \left ( 2~-~ \bar{Q}' \right )} ~
		  \left ( \frac{d m}{d \tau} \right ) _{out} ~c^{2} ~,
\nonumber \\
\frac{d \vec{p}_{out}}{d \tau} &=& -~ \gamma_{c} ~
		  \sqrt{\bar{Q}' \left ( 2~-~ \bar{Q}' \right )} ~
		  \left ( \frac{d m}{d \tau} \right ) _{out} ~\vec{v}_{c} ~,
\nonumber \\
\vec{v}_{c} &=& \frac{\bar{Q}' ~\gamma~\vec{v} ~+~
		\left ( 1 ~-~ \bar{Q}' \right ) ~c ~\vec{e}_{1} / w_{1} }{
		\bar{Q}' ~\gamma ~+~ \left ( 1 ~-~ \bar{Q}' \right ) / w_{1} } ~,
\nonumber \\
\bar{Q}' &\equiv& \frac{\bar{Q}'_{pr, 1}}{\bar{Q}'_{ext}} ~.
\end{eqnarray}
The relations for $E_{out}$ and $\vec{p}_{out}$ seem to be only a little
modified standard relations (invariant multiplier
$\sqrt{\bar{Q}' ( 2~-~ \bar{Q}')}$ is present).
However, this is not true: the velocity is $\vec{v}_{c}$ and it is not equal
to the velocity of the particle $\vec{v}$ (except for the case
$\bar{Q}'$ $=$ 1): Eqs. (69) hold. Also, comparison
of Eqs. (87)-(88) with Eq. (80) yields
\begin{eqnarray}\label{90}
\frac{d M_{out}}{d \tau} &=& -~
	    \sqrt{ \frac{\bar{Q}'_{pr, 1}}{\bar{Q}'_{ext}} ~ \left ( 2 ~-~
	    \frac{\bar{Q}'_{pr, 1}}{\bar{Q}'_{ext}} \right )} ~
	    \left ( \frac{d m}{d \tau} \right ) _{out} ~.
\end{eqnarray}
Eqs. (69), (89)-(90) state that the  formulation ''mass equals
energy divided by $c^{2}$'' holds for radiation ($dE_{out}$, $dM_{out}$),
but not for the mass of the particle ($dE_{out}$, $dm$), if $\bar{Q}'$ $\ne$ 1.
It is worth mentioning that the special case
$\bar{Q}'_{pr, 1}$ $/$ $\bar{Q}'_{ext}$ $=$ 1,
yielding $\vec{p}'_{o}$ $=$ 0 (generally believed to be the case of perfect
absorption within geometrical optics approximation and which was treated by
Robertson 1937, see also Robertson and Noonan 1968, p. 114),
leads to very simple physics: $\vec{v}_{c}$ $=$ $\vec{v}$, according
to Eq. (89), $d M_{out}$ $/$ $d \tau$ $=$ $-$
$(d m / d \tau)_{out}$, according to Eq. (90), and $(d E)_{out}$ $=$ $-$
$\gamma$ $(d m)_{out}$ $c^{2}$.

If the particle is not irradiated, then Eq. (41) holds:
$(d m / d \tau)_{e}$ $=$ $-$ $E'_{o}$ $/$ $c^{2}$. Comparison with Eq. (83)
yields: $(d m / d \tau)_{e}$ $=$ $-$ $(d M_{out} / d \tau)_{e}$.

\subsubsection{Mirror-like spherical surface and geometrical optics -- specular
reflection}
On the basis of Eq. (79) we know that
0 $<$ $\bar{Q}'_{pr, 1}$ $/$ $\bar{Q}'_{ext}$ $\le$ 2. Moreover, one could say
that the special case of perfect absorption within geometrical optics
approximation would yield $\bar{Q}'_{pr, 1}$ $/$ $\bar{Q}'_{ext}$ $=$ 1,
according to Eq. (73) and according to the conventional statement that the
absorbed radiation is isotropically reemitted (in the particle's
frame of reference) -- $\vec{p}_{o} '$ $=$ $0$.
Moreover, a statement that $\bar{Q}'_{pr, 1}$ $/$ $\bar{Q}'_{ext}$ $=$ 2
should hold for spherical particle with mirror-like surface (totally reflecting
sphere), can also be found. However, the statements presented in the
last two sentences are not correct. In order to prove this conclusion, we will
treat a nonrotating totally reflecting sphere, within  a
geometrical optics approximation. We will consider specular reflection of the
electromagnetic radiation at the surface of the spherical
particle, in the particle's rest frame of reference (for the sake of
brevity, we will not use primed quantities, in this subsection).

Let the center of the cartesian reference frame $x-y-z$ be situated at the
center of the spherical particle and the incoming beam of photons is
characterized by unit vector of momentum $\hat{\vec{p}}_{inc}$ $=$ $-$
$\hat{\vec{x}}$ $=$ $-$ ( 1,~ 0, ~0 ). Any point
on the surface of the particle, which interacts with the incoming radiation,
can be characterized by coordinates of the surface position vector in the
following way:
\begin{eqnarray}\label{91}
\vec{R} &=& R ~ \vec{n} ~,
\nonumber \\
\vec{n} &=& ( \cos \alpha ~ \cos \delta , ~
	       \sin \alpha ~ \cos \delta , ~ \sin \delta ) ~,
\nonumber \\
\alpha &\in& \langle  - ~ \frac{\pi}{2} , ~+ ~\frac{\pi}{2} \rangle ~,
\nonumber \\
\delta &\in& \langle  - ~ \frac{\pi}{2} , ~+ ~\frac{\pi}{2} \rangle ~,
\end{eqnarray}
where $R$ is particle's radius and $\vec{n}$ is unit vector.

Now, we
have to use {\it the law of reflection}: {\it The angle of incidence equals
the angle of reflection, and the incident and reflected rays are in the
same plane.} The angle of incidence $\varphi$ is given by condition
\begin{equation}\label{92}
\cos \varphi = | ~ \vec{n} \cdot ( -~ \hat{\vec{x}} ) ~| =
	       \cos \alpha ~ \cos \delta ~.
\end{equation}
The plane of incidence is characterized by its normal unit vector $\vec{N}$:
\begin{eqnarray}\label{93}
\vec{N} &=& \frac{ \left ( -~ \hat{\vec{x}} \right ) \times \vec{n}}{
	      \left | \hat{\vec{x}}  \times \vec{n} \right |} =
\nonumber \\
	&=& \frac{( 0, ~ \sin \delta, ~ - \sin \alpha ~ \cos \delta )}{
	    \sqrt{\sin^{2} \delta ~+~ \sin^{2} \alpha ~ \cos^{2} \delta}} ~.
\end{eqnarray}
The equation of the plane of incidence is given as $\vec{N}$ $\cdot$ $\vec{R}$
$+$ $k$ $=$ 0, which yields $k$ $=$ $-$ $\vec{N}$ $\cdot$ $\vec{R}$ $=$ 0,
according to Eqs. (91) and (93). The plane of incidence is identical to the
plane of reflection:
\begin{eqnarray}\label{94}
\vec{N} \cdot \hat{\vec{p}}_{out} &=& 0 ~,
\nonumber \\
\hat{\vec{p}}_{out} \cdot \vec{n} &=& \cos \varphi ~,
\end{eqnarray}
where also the condition ''the angle of incidence equals the angle of
reflection'' was added (the case $\alpha$ $=$ 0, $\delta$ $=$ 0 yields
$\hat{\vec{p}}_{out}$ $=$ $+$ $\hat{\vec{x}}$). Conditions formulated
by Eq. (94) yield, on the basis of Eqs. (91)-(93):
\begin{eqnarray}\label{95}
\hat{\vec{p}}_{out} &\equiv& (x_{out}, ~y_{out}, ~z_{out}) ~,
\nonumber \\
x_{out} &=& ( \cos^{2} \alpha ~-~ \sin^{2} \alpha ) ~\cos^{2} \delta ~-~
			    \sin^{2} \delta ~,
\nonumber \\
y_{out} &=& \left ( 1~-~ x_{out} \right ) ~
      \frac{\cos \alpha ~\sin \alpha ~\cos^{2} \delta}{
      \sin^{2} \alpha ~\cos^{2} \delta ~+~ \sin^{2} \delta} ~,
\nonumber \\
z_{out} &=& \left ( 1~-~ x_{out} \right ) ~
      \frac{\cos \alpha ~\cos \delta ~\sin \delta}{
      \sin^{2} \alpha ~\cos^{2} \delta ~+~ \sin^{2} \delta} ~.
\end{eqnarray}

Unit vector along the small circle parallel with the equator $\vec{e}_{\alpha}$,
and, unit vector along the great circle normal to the equator $\vec{e}_{\delta}$,
both localized at the position vector $\vec{R}$, are given by formulae
\begin{eqnarray}\label{96}
\vec{e}_{\alpha} &=& \left ( \cos \left ( \alpha + \frac{\pi}{2} \right ) , ~
		     \sin \left ( \alpha + \frac{\pi}{2} \right ) , ~0 \right )
\nonumber \\
       &=& ( - ~ \sin \alpha , ~\cos \alpha , ~0 ) ~,
\nonumber \\
\vec{e}_{\delta} &=& \vec{n} \times \vec{e}_{\alpha}
\nonumber \\
       &=& ( - ~ \cos \alpha ~ \sin \delta , ~-~ \sin \alpha ~ \sin \delta,
	   ~ \cos \delta ) ~,
\end{eqnarray}
since the relation for $\vec{e}_{\delta}$ is a consequence of the relation
$\vec{e}_{\alpha}$ $\times$ $\vec{e}_{\delta}$ $=$ $\vec{n}$.
Element of the length vector $\vec{dl}$ $\equiv$ $dx$ $\hat{\vec{x}}$ $+$
$dy$ $\hat{\vec{y}}$ $+$ $dz$ $\hat{\vec{z}}$ $=$ $R$ $d \vec{n}$ is, on the
basis of Eq. (91):
\begin{eqnarray}\label{97}
\vec{dl} &=& R \left ( -~ \sin \alpha ~ \cos \delta ~ d \alpha ~-~
		       \cos \alpha ~ \sin \delta~ d \delta \right ) ~
	       \hat{\vec{x}}
\nonumber \\
   & & +~R \left ( \cos \alpha ~ \cos \delta ~ d \alpha ~-~
		       \sin \alpha ~ \sin \delta~ d \delta \right ) ~
	       \hat{\vec{y}}
\nonumber \\
   & & +~R \cos \delta ~ d \delta ~\hat{\vec{z}} ~.
\end{eqnarray}
Components of the length vector along the vectors $\vec{e}_{\alpha}$ and
$\vec{e}_{\delta}$ are:
\begin{eqnarray}\label{98}
dl_{\alpha} &=& \vec{dl} \cdot \vec{e}_{\alpha} =
		R ~ \cos \delta ~ d \alpha ~,
\nonumber \\
dl_{\delta} &=& \vec{dl} \cdot \vec{e}_{\delta} =
		R ~ d \delta ~.
\end{eqnarray}
Vector of the elementary/infinitesimal area is
\begin{eqnarray}\label{99}
d \vec{A} &=& \left ( dl_{\alpha} ~ \vec{e}_{\alpha} \right ) ~\times ~
	      \left ( dl_{\delta} ~ \vec{e}_{\delta} \right ) =
	      R^{2} ~ \cos \delta ~ d \alpha ~d \delta ~ \vec{n}
\nonumber \\
	  &=& R^{2} ~ d \Omega ~ \vec{n} ~,
\end{eqnarray}
where $d \Omega$ is the solid angle element at the position ($\alpha$, $\delta$).
The amount of energy (per unit time) hitting the area $d A$ at the position
$\vec{R}$ is $S$ $d A_{\bot}$, where $S$ is the flux density
of radiation energy (energy flow through unit area perpendicular to the ray of
photons per unit time) and
\begin{eqnarray}\label{100}
d A_{\bot} &=& d \vec{A} ~ \cdot ~\hat{\vec{x}} =
	     ( R^{2} ~d \Omega ) ~ \cos \alpha ~\cos \delta =
\nonumber \\
     &=&     R^{2} ~ \cos \alpha ~
	     \cos^{2} \delta ~d \alpha ~ d \delta ~,
\end{eqnarray}
if also Eq. (91) is used. The quantity $d A_{\bot}$ is projection
of the area $d A$ on the direction normal to $\hat{\vec{x}}$.

On the basis of Eqs. (91) and (100) we can write for the incoming momentum
per unit time
\begin{eqnarray}\label{101}
\frac{d \vec{p}_{in}}{d t} &=& \frac{1}{c} ~ S ~ \int d A_{\bot} ~
			       \hat{\vec{p}}_{inc} =
\nonumber \\
 &=&  \frac{S}{c} ~ \hat{\vec{p}}_{inc} ~ \int_{- \pi / 2}^{\pi / 2} d \alpha ~
      \int_{- \pi / 2}^{\pi / 2} d \delta ~\left \{ R^{2} ~ \cos \alpha ~
      \cos^{2} \delta \right \} =
\nonumber \\
&=&  \frac{S}{c} ~ \pi ~ R^{2} ~\hat{\vec{p}}_{inc} ~.
\end{eqnarray}
The outgoing momentum per unit time is, on the basis of Eqs. (95) and (100),
\begin{equation}\label{102}
\frac{d \vec{p}_{out}}{d t} = \frac{1}{c} ~ S ~ \int d A_{\bot} ~
			       \hat{\vec{p}}_{out} = 0 ~.
\end{equation}
The mirror-like spherical particle obtains the following momentum per unit time:
\begin{equation}\label{103}
\frac{d \vec{p}}{d t} = \frac{d \vec{p}_{in}}{d t} ~-~
\frac{d \vec{p}_{out}}{d t} = \frac{S}{c} ~ \pi ~ R^{2} ~\hat{\vec{p}}_{inc} ~,
\end{equation}
which immediately follows from Eqs. (101)-(102). Comparison with Eq. (70)
or (72) yields $\bar{Q}'_{pr, 1} =$ 1. This is consistent with the results
presented in van de Hulst (van de Hulst 1981, p. 161: see Table 13
"Efficiency Factors for Totally Reflecting Spheres"). However, comparison of
Eqs. (73) and (102) would yield $\bar{Q}'_{pr, 1} / \bar{Q}'_{ext}$ $=$ 1, and,
thus, $\bar{Q}'_{ext}$ $=$ 1. But this is not consistent with the results
of the Mie theory presented in van de Hulst (1981, p. 161 -- Table 13).
This seems surprisingly,
since van de Hulst states on p. 223: "a smooth totally reflecting sphere with
radius large compared to the wavelength scatters light by reflection
isotropically". Where's the problem? In order to understand physics of the
results, we have to realize that the result presented by Eq. (102) is based on
geometrical optics and the result does not respect all physical interactions, only
reflection is considered. In reality, "extinction paradox" exists (van de Hulst
1981, p. 107) and diffracted light plays a non-negligible role in treating
the incoming and outgoing radiation, even for large particles (in comparison
with a wavelength of the interacting light);
"diffraction $=$  small-angle scattering" (van de Hulst 1981, p. 107).
We know that "the diffracted light gives a zero contribution" to the radiation
pressure of large spheres (van de Hulst 1981, p. 225), but diffraction cannot
be neglected in a separate treatement of the incoming and outgoing radiation.
In order to be consistent with equations of the type of Eq. (52), i. e.,
Eqs. (53) and (78)-(79), we have to take into account diffraction, also. Thus,
we have $\bar{Q}'_{pr, 1} =$ 1 and $\bar{Q}'_{ext}$ $=$ 2, both for perfectly
absorbing sphere ($\bar{Q}'_{sca}$ $=$ 1 due to the diffracted light and
$\bar{Q}'_{abs}$ $=$ 1) and for totally reflecting sphere
($\bar{Q}'_{sca}$ $=$ 1 $+$ 1, 1 due to the diffracted light $+$ 1
due to the reflected light, $\bar{Q}'_{abs}$ $=$ 0):
\begin{equation}\label{104}
\frac{\bar{Q}'_{pr, 1}}{\bar{Q}'_{ext}} = \frac{1}{2} ~.
\end{equation}
The result $\bar{Q}'_{pr, 1}$ $/$ $\bar{Q}'_{ext}$ $=$ 1/2 holds for spherical
particles not only with perfect absorption, but, also with mirror-like
surface characterized by specular reflection. Eq. (73) holds for the
incoming and outgoing radiation. Eq. (78) yields
\begin{eqnarray}\label{105}
d M_{out} &=& \sqrt{3} ~\frac{w_{1}^{2}~S~A'}{c^{2}} ~ d \tau ~.
\end{eqnarray}
One would await, on the basis of a simple idea about the isotropic light
reemission or scattering (diffraction is neglected), the result without
the numerical factor $\sqrt{3}$ -- the same result would be obtained on the
basis of the following frequent idea (see also Sec. 8):
"Photons at rest do not exist! It is therefore somehow artificial to speak
of the photon rest mass. It is more logical to use the relation
$m$ $=$ $h$ $\nu$ $/$ $c^{2}$ for the definition of the photon mass."
(Demtr\H{o}der 2006, p. 92).

\subsubsection{Summary}
We have already mentioned "the extinction paradox" (van de Hulst 1981, p. 107).
It states that "a large particle removes from the incident beam exactly
{\it twice} the amount of light it can intercept" (van de Hulst 1981, p. 107).
As van de Hulst explains, the explanation is in the fact that "({\it a}) {\it all}
scattered light, including that at small angles, is counted as removed
from the beam, and ({\it b}) that the observation is made at a very great
distance, i. e., far beyond the zone where a shadow can be distinguished".

Let us take into account the values presented in Table 13 in van de Hulst (1981,
p. 161). One can immediately realize that the values $Q_{pr}$ greater
than 2 would be in contradiction with equations analogous to Eqs. (78)-(79),
if $\vec{p}_{o}'$ $=$ ( 1 $-$ $Q_{pr} '$ ) $\vec{p}_{i}'$ would be the
condition for the Poynting-Robertson effect. Thus, the case of
"perfectly absorbing case" does not correspond to $\vec{p}_{o}'$ $=$ 0
within geometrical optics approximation plus diffraction (relevant
physical processes), i. e., isotropic reradiation/reemission. Instead of this
we have to use Eq. (11) and the "perfectly absorbing case" yields
$\vec{p}_{o}'$ $=$ ( 1 $-$ $Q_{pr} '$ $/$ $Q_{ext} '$ ) $\vec{p}_{i}'$ $=$
(1 $/$ 2) $\vec{p}_{i}'$ for large particles
[$2$ $\pi$ $\times$ radius of the particle / (wavelength of the incident
radiation) $\gg$ 1]. The term "isotropic reemission" does not hold,
strictly speaking, neither for small particle, nor for large sphere.
Similar situation holds for totally reflecting sphere.
[Although, within geometrical optics approximation, the term "isotropic
reradiation" is used, which is based on the fact that "for large spheres
we may separate the effects of geometrical optics (reflection and refraction)
and the effect of diffraction" (van de Hulst 1981, p. 225).]

We have found that the condition for the validity of the
Poynting-Robertson effect is given by Eq. (73). The case of "perfectly
absorbing spherical particle" is given by Eq. (104), or, by the condition
$\vec{p}'_{o}$ $=$ 0.5 $\vec{p}'_{i}$ for large spherical particle.
This condition differs from the
condition used by Robertson (1937): it is generally accepted that the
Robertson's case corresponds to $\vec{p}'_{o}$ $=$ 0. E. g.: "Consider
a spherical particle with proper mass $m$ and world-velocity $u^{\mu}$
which scatters (whether by reflection or by absorption and re-radiation)
electromagnetic radiation isotropically in all directions relative to its
proper system. ... The proper system $\bar{x}^{\alpha}$ of the particle
at a given event is used for writing down the equation of motion. The
force on the particle is then due only to the incident radiation, because
the re-emitted radiation is isotropic." (Robertson and Noonan 1968, p. 114).
Similarily, in his original paper, Robertson writes on p. 427:
"the particle is assumed to possess spherical symmetry and to radiate
isotropically in the rest system" of the particle. On the basis of this
assumption, the author deals with geometrical optics approximation
and he writes equation of motion in the form corresponding to our
equation $dp^{\mu}$ $/$ $d \tau$ $=$ $dp_{in}^{\mu}$ $/$ $d \tau$ $-$
$dp_{out}^{\mu}$ $/$ $d \tau$, where $dp_{in}^{\mu}$ $/$ $d \tau$ is
given by Eq. (75) and $dp_{out}^{\mu}$ $/$ $d \tau$ by Eq. (76).
However, Robertson asserts that $\bar{Q}'_{pr, 1}$ $=$ 1 and
$\bar{Q}'_{ext}$ $=$ 1 (see Eqs. 2.7 and 2.8 on p. 427 in Robertson 1937,
and, also the statement that "the first term
[$( w_{1}^{2} S A' / c )~ b_{1}^{\mu}$] represents the total power-force
four-vector due to the incident radiation, with which it coincides
completely in direction, and the second
[$( w_{1}^{2} S A' / c )~ u^{\mu} / c$] that due to the radiation re-emitted
at the rate [$w_{1}^{2} S A' c$] "). The statements of Robertson are not
correct, since $\bar{Q}'_{pr, 1}$ $=$ 1 and
$\bar{Q}'_{ext}$ $=$ 2, in reality (see Eq. 104). The Robertson's
approach would yield, on the basis of equations similar to Eqs. (77)-(78)
(with the transformation $\bar{Q}'_{pr, 1}$ $/$ $\bar{Q}'_{ext}$
$\longrightarrow$ $\bar{Q}'_{pr, 1}$, see e. g., Eq. 122 in Kla\v{c}ka 1992),
that 0 $<$ $\bar{Q}'_{pr, 1}$ $<$ 2, which would be an inconsistency
between the Mie theory and the relativity theory. These considerations
of Robertson are equivalent to that of Poynting
(1903, p. 539: "... it is receiving a stream
of momentum $w_{1}^{2} S A' / c$ directed from the sun. Its own radiation
outwards being equal in all directions has zero resultant pressure." ... we
have used symbols of our paper). In reality, the case
$\vec{p}'_{o}$ $=$ 0, considered by Poynting (1903), Robertson (1937),
Wyatt and Whipple (1950, p. 558: "the process of absorption and re-emission
produces no net force on a particle when one chooses to work with a
stationary frame referred to the particle" ... the part "absorption and"
should be omitted in the formulation) and others, does not
hold and the correct form is $\vec{p}'_{o}$ $=$ 0.5 $\vec{p}'_{i}$. Thus,
we have understand the real physics of the P-R effect for the first time, now!
We are aware that the effect of diffraction must not be neglected, even in the
case of geometrical optics approximation.

Finally, let us consider a large plane mirror of geometrical area $A'$
characterized by specular reflection (totally reflecting plane)
and normal incident radiation. We have
$C_{abs} '$ $=$ 0, $C_{sca} '$ $=$ 2 $A'$, $C_{ext} '$ $=$ 2 $A'$ and
$\langle \cos \theta ' \rangle$ $=$ 0, $C_{pr, 1} '$ $=$ 2 $A'$. We can write \\
$\vec{p}_{o}'$ $=$ ( $1 -$ $C_{pr, 1} '$ $/$ $C_{ext} '$) $\vec{p}_{i}'$ $=$ 0
... ($**$),  \\
and Eqs. (11) and ($**$) yield \\
$E_{i} '$ $=$ $S'$ $C_{ext} '$ $=$ 2 $S'$ $A'$,
$\vec{p}_{i}'$ $=$ ( $E_{i} '$ $/$ $c$ ) $\vec{e}_{1}'$, \\
$E_{o} '$ $=$ $E_{i} '$ $=$ 2 $S'$ $A'$, $\vec{p}_{o}'$ $=$ 0. \\
Eqs. (50) and (53) yield, then: \\
$dM_{in}$ $=$ 0, $dM_{out}$ $=$ (2 $S'$ $A'$ $/$ $c^{2}$ ) $d \tau$. \\
On the basis of geometrical optics considerations without the effect of
diffraction ($C_{pr, 1} '$ $=$ 2 $A'$, $C_{ext} '$ $=$ $A'$ would have to be
used in Eqs. 50 and 54), one would await that
$\vec{p}_{o}'$ $=$ $-$ $\vec{p}_{i}'$ and ( $dM_{out}$ )$_{naive}$ $=$
( $dM_{in}$ )$_{naive}$ $=$ 0 -- the idea that photon's mass equals its energy
divided by $c^{2}$ would yield ( $dM_{in}$ )$_{ph}$ $=$ ( $dM_{out}$ )$_{ph}$
$=$ ( $S'$ $A'$ $/$ $c^{2}$ ) $d \tau$.
The result for $dM_{out}$ can be easily
physically understood: the situation is similar to finding a mass
of two photons, each of which travels in the same direction but
orientations of their motions are different --	reflected and
diffracted photons are important, in our case.

\subsection{Orbital evolution}
In general, equation of motion for arbitrarily shaped particle is given by
Eqs. (38) or (39), or, by Eqs. (40) or (41). Laboratory experiments
carried out by Krauss and Wurm (2004) proved that nonspherical particles
behave in a different way than spherical ones.

The Poynting-Robertson effect yields the following property, as for the
particle's orbital evolution in gravitational and electromagnetic fields of a
central star: semi-major axis and eccentricity secularly decrease
(see Kla\v{c}ka 2004: Secs. 6.1 and 6.2 - mainly 6.2.3, for more details),
if the optical properties of the particle do not change
(an increase of the orbital elements can occur, if $\beta$ increases:
see Eq. 12 in Kla\v{c}ka 1993).
This is fulfilled under assumptions presented in
Sec. 7.3, i.e. $\bar{C}'_{pr, 2}$ $=$ $\bar{C}'_{pr, 3}$ $=$
$\vec{F}'_{e}$ $=$ 0 and Eqs. (73) or (74) hold. The equation of motion
is given by Eqs. (70) or (71), then. As a consequence, the equation
of motion holds for particle with spherically symmetric mass
distribution.

As for presentations in current literature, the statements are not
consistent with the discussed results.

\subsubsection{Accelerations of the spherical particle}
Relations for the accelerations of the spherical particle interacting with
the incident electromagnetic radiation can be obtained from Eqs. (75)-(76).
The relevant relations for the accelerations
can be summarized in the following way.

Eq. (75) yields for the four-acceleration of the spherical particle
due to the incoming radiation:
\begin{eqnarray}\label{106}
\left ( \frac{d u^{\mu}}{d \tau} \right )_{in} &=& X~
	     \left ( b_{1}^{\mu} - \frac{u^{\mu}}{c} \right ) ~,
\nonumber \\
X &\equiv& \frac{w_{1}^{2} ~ S ~A' ~\bar{Q}'_{ext}}{m~c} ~,
\end{eqnarray}
or, to the first order in $\vec{v} / {c}$
\begin{equation}\label{107}
\left ( \frac{d \vec{v}}{d t} \right )_{in} =
	   \frac{S ~A' ~\bar{Q}'_{ext}}{m~c} ~
	   \left \{ \left ( 1 ~-~ \frac{\vec{v} \cdot \vec{e}_{1}}{c} \right )
	  ~ \vec{e}_{1} - \frac{\vec{v}}{c} \right \} ~.
\end{equation}
The change of the particle's mass is
\begin{equation}\label{108}
\left ( \frac{d m}{d \tau} \right )_{in} =
	   \frac{w_{1}^{2} ~ S ~A' ~\bar{Q}'_{ext}}{c^{2}} ~.
\end{equation}

Eq. (76) enables to find the four-acceleration of the spherical particle
due to the outgoing radiation.
Eq. (76) yields for the force acting on the particle
$( d p^{\mu} / d \tau )_{out}$ $=$ $-$ $d p_{out}^{\mu} / d \tau$, which yields for
the four-acceleration of the particle
\begin{eqnarray}\label{109}
\left ( \frac{d u^{\mu}}{d \tau} \right )_{out} &=& -~ X~
	     \left ( 1 ~-~ \frac{\bar{Q} '_{pr, 1}}{\bar{Q}' _{ext}} \right ) ~
	     \left ( b_{1}^{\mu} - \frac{u^{\mu}}{c} \right ) ~,
\nonumber \\
X &\equiv& \frac{w_{1}^{2} ~ S ~A' ~\bar{Q}'_{ext}}{m~c} ~,
\end{eqnarray}
or, to the first order in $\vec{v} / {c}$
\begin{eqnarray}\label{110}
\left ( \frac{d \vec{v}}{d t} \right )_{out} &=& -~
	     \frac{S ~A' ~\bar{Q}'_{ext}}{m~c} ~
	     \left ( 1 ~-~ \frac{\bar{Q} '_{pr, 1}}{\bar{Q}' _{ext}} \right )
\nonumber \\
& & \times \left \{ \left ( 1 ~-~ \frac{\vec{v} \cdot \vec{e}_{1}}{c} \right )
	   ~ \vec{e}_{1} - \frac{\vec{v}}{c} \right \} ~.
\end{eqnarray}
The change of the particle's mass is
\begin{equation}\label{111}
\left ( \frac{d m}{d \tau} \right )_{out} = -~
	   \frac{w_{1}^{2} ~ S ~A' ~\bar{Q}'_{ext}}{c^{2}} ~.
\end{equation}

Total four-acceleration of the spherical particle is
\begin{eqnarray}\label{112}
\left ( \frac{d u^{\mu}}{d \tau} \right )_{particle} &=&
	      \left ( \frac{d u^{\mu}}{d \tau} \right )_{in} ~+~
	      \left ( \frac{d u^{\mu}}{d \tau} \right )_{out}
\nonumber \\
&=& \frac{\bar{Q} '_{pr, 1}}{\bar{Q}' _{ext}} ~X~
    \left ( b_{1}^{\mu} - \frac{u^{\mu}}{c} \right ) ~,
\nonumber \\
X &\equiv& \frac{w_{1}^{2} ~ S ~A' ~\bar{Q}'_{ext}}{m~c}
\end{eqnarray}
and the mass of the particle is conserved:
$( d m / d \tau )_{particle}$ $=$ $( d m / d \tau )_{in}$ $+$
$( d m / d \tau )_{out}$ $=$ 0. To the first order in $\vec{v} / {c}$
\begin{eqnarray}\label{113}
\left ( \frac{d \vec{v}}{d t} \right )_{particle} &=&
      \frac{\bar{Q} '_{pr, 1}}{\bar{Q}' _{ext}} ~
      \frac{S ~A' ~\bar{Q}'_{ext}}{m~c} ~
      \left \{ \left ( 1 ~-~ \frac{\vec{v} \cdot \vec{e}_{1}}{c} \right )
	   ~ \vec{e}_{1} - \frac{\vec{v}}{c} \right \} ~.
\end{eqnarray}

If $\bar{Q} '_{pr, 1} / \bar{Q}' _{ext}$ $=$ 1,
then $( d \vec{v} / d t )_{out}$ $=$ 0 (see Eq. 110). Further,
0 $<$ $\bar{Q} '_{pr, 1} / \bar{Q}' _{ext}$ $<$ 2, according to Eq. (79).
If $\bar{Q} '_{pr, 1} / \bar{Q}' _{ext}$ $<$ 1, then the re-radiated
light yields $( d \vec{v} / d t )_{out}$ $\propto$ $+$ $\vec{v}$.
If $\bar{Q} '_{pr, 1} / \bar{Q}' _{ext}$ $>$ 1, then the re-radiated
light yields $(d \vec{v} / d t)_{out}$ $\propto$ $-$ $\vec{v}$. The case
$(d \vec{v} / d t)_{out}$ $\propto$ $-$ $\vec{v}$ leads to a decrease of particle's
total energy (kinetic plus potential) and to an increase of the orbital speed
$| \vec{v} |$, if the optical properties of the particle do not significantly
change during the particle's motion around a star.

\subsubsection{Karttunen {\it et al.} (2007)}
Karttunen {\it et al.} (2007, p. 201) state: ''When a small body absorbs
and emits radiation, it loses its orbital angular momentum and the body
spirals to the Sun.''

The statement holds only for spherical particle
orbiting a star (Sun), under the assumption that the particle is characterized
with spherically symmetric mass distribution and constant optical properties.
Moreover, even when spherical particle does not
absorb radiation, it may (secularly) lose orbital angular momentum (see also
Sec. 7.3.3 as a very special case, or any elastic scattering). If optical
parameters of the particle change, then the particle may not lose its
(secular) orbital angular momentum.
If the particle is nonspherical, there may not exist systematic secular
decrease of particle's semi-major axis and eccentricity, and, moreover,
particle's orbital plane may change.

\subsubsection{Carroll and Ostlie (2007)}
Carroll and Ostlie (2007, p. 806) describe the Poynting-Robertson effect
in this way: ''When particles absorb sunlight, they must re-radiate that
energy again if they are to remain in thermal equilibrium. The original light
was emitted from the Sun isotropically, but in the Sun's rest frame the
re-radiated light is concentrated in the direction of motion of the particle.
Since the re-radiated light carries away momentum as well as energy, the
particle slows down and its orbit decay.''

The statement does not hold for arbitrarily shaped particles. It holds,
partially, only for dust grains with spherically symmetric mass distribution
and constant optical properties. If such type of spherical particles is
considered, then Eq. (74) holds for the outgoing radiation. Eq. (74) shows that
''re-radiation'' (but also scattering, in general) occurs both in directions
$d\vec{p}_{in}/dt$ $\propto$ $\vec{e}_{1}$ and $\vec{v}$. We can write,
on the basis of Eqs. (77) and (78) (to the first order in $\vec{v}/c$):
$d \vec{p}_{out} / dt$ $=$ ( $S$ $A'$ $\bar{Q}_{ext}'$ $/$ $c$ ) $\times$
[ ( 1 $-$ $\bar{Q}'$ ) ( 1 $-$  $\vec{v} \cdot \vec{e}_{1}$ $/$ $c$ ) ~
$\vec{e}_{1}$ $+$ $\bar{Q}'$ $\vec{v}$ $/$ $c$ ] $=$
( $S$ $A'$ $\bar{Q}_{ext}'$ $/$ $c$ ) $\times$
[ ( 1 $-$  $\vec{v} \cdot \vec{e}_{1}$ $/$ $c$ ) ~ $\vec{e}_{1}$ $-$
$\bar{Q}'$ $\vec{e}_{1}$ $+$
$\bar{Q}'$ ( ( $\vec{v} \cdot \vec{e}_{1}$ $/$ $c$ ) ~
$\vec{e}_{1}$ $+$ $\vec{v}$ $/$ $c$ ) ],
$\bar{Q}'$ $\equiv$ $\bar{Q}_{pr, 1}'$ $/$ $\bar{Q}_{ext}'$. Only the
direction $\vec{v}$ is considered in the explanation of Carroll and Ostlie
('' ... in the Sun's rest frame the re-radiated light is concentrated
in the direction of motion of the particle.'').

The statement that ''the re-radiated light carries away momentum as well as
energy'' is correct. Mathematical description of this physical process
is given by Eq. (76). But the complete sentence
''Since the re-radiated light carries away momentum as well as energy, the
particle slows down and its orbit decay.'' is physically incorrect, in general.
Really, Eq. (76) yields for the force acting on the particle
$( d p^{\mu} / d \tau )_{out}$ $=$ $-$ $d p_{out}^{\mu} / d \tau$, which yields for
the four-acceleration of the particle results given by Eqs. (109)-(110).
If $\bar{Q} '_{pr, 1} / \bar{Q}' _{ext}$ $=$ 1,
then $( d \vec{v} / d t )_{out}$ $=$ 0 (see Eq. 110)
and the particle does not slow down.
If $\bar{Q} '_{pr, 1} / \bar{Q}' _{ext}$ $<$ 1, then the re-radiated
light yields $(d \vec{v} / d t)_{out}$ $\propto$ $+$ $\vec{v}$.
If $\bar{Q} '_{pr, 1} / \bar{Q}' _{ext}$ $>$ 1, then the re-radiated
light yields $(d \vec{v} / d t)_{out}$ $\propto$ $-$ $\vec{v}$.
The case of "perfectly absorbing sphere", treated by Robertson (1937),
corresponds to $\bar{Q} '_{pr, 1} / \bar{Q}' _{ext}$ $=$ 1/2
within geometrical optics approximation, and, thus,
the re-radiated light yields $(d \vec{v} / d t)_{out}$ $\propto$ $+$ $\vec{v}$,
according to Eq. (110).

\subsubsection{Lissauer and Murray (2007)}
Lissauer and Murray (2007, p. 806) explain:

"A small particle in orbit
around the Sun absorbs solar radiation and reradiates the energy isotropically
in its own frame. The particle thereby preferentially radiates (and loses
momentum) in the forward direction in the inertial frame of the Sun. This
leads to a decrease in the particle's energy and angular momentum and causes
dust in bound orbits to spiral sunward. This effect is called the
Poynting-Robertson drag.

The net force on a rapidly rotating dust grains is given by \\
$\vec{F}_{rad} \approx [ ( L~ Q_{pr} ~A ) / (4 ~\pi ~c ~r^{2} ) ]$
[ $ ( 1 ~-~ 2~v_{r} / c ) ~\hat{\vec{r}} ~-~ ( v_{\Theta} / c ) ~
\hat{\vec{\Theta}}$ ]~.~~~   (50-LM) \\
The first term in Eq. (50-LM) is that due to radiation pressure and the
second and the third terms (those involving the velocity of the particle)
represent the Poynting-Robertson drag.

From this discussion, it is clear that small-sized dust grains in the
interplanetary medium are removed: (sub)-micronsized grains are blown out
of the solar system, whereas larger particles spiral inward toward the Sun."

(Explanation to symbols in Eq. 50-LM are given also on p. 806 of the paper:
"$A$ is the particle's geometric cross section, $L$ is the solar luminosity,
$c$ is the speed of light, $r$ is the heliocentric distance, and
$Q_{pr}$ is the radiation pressure coefficient, which is equal to unity
for a perfectly absorbing particle and is of order unity unless the particle
is small compared to the wavelength of the radiation.")

Now, we will treat the real physics.

i) \\
"A small particle ... ."
(Lissauer and Murray 2007, p. 806)

The equation of motion corresponding to the Poynting-Robertson effect
holds for (nonrotating) spherical particles.

ii) \\
"A small particle ... reradiates the energy isotropically in its own frame."
(Lissauer and Murray 2007, p. 806)

General assumption/condition for the validity of the Poynting-Robertson effect
is formulated by Eq. (73). This equation shows that $\vec{p}_{o} '=$ 0
if and only if $\bar{Q}'_{pr, 1} / \bar{Q}'_{ext}$ $=$ 1.  Thus, the statement
of the authors, presented above, corresponds only to a special combination
of optical properties of the spherical particle and the wavelength of the
incident electromagnetic radiation. The authors probably assume perfectly
absorbing spherical particles within geometrical optics approximation
(moreover, diffraction is ignored by the authors), but this case is characterized
by the condition $\bar{Q}'_{pr, 1} / \bar{Q}'_{ext}$ $=$ 1/2, as it is given by
Eq. (104), and $\vec{p}_{o} '$ $=$ (1/2) $\vec{p}_{i} '$.

iii) \\
"The particle therefore preferentially radiates in the forward direction
in the inertial frame of the Sun." (Lissauer and Murray 2007, p. 806)

According to Eq. (76), we can write for the outgoing radiation for the
P-R effect (first order in $\vec{v} / c$ is considered): \\
$d \vec{p}_{out} / d t$ $=$ ( $S ~A' ~\bar{Q}'_{ext} / c$ )
$\left \{ ( 1 ~-~\bar{Q} '_{pr, 1} / \bar{Q}' _{ext} ) ( 1 ~-~
\vec{v} \cdot \vec{e}_{1} / c ) ~\vec{e}_{1} ~+~
( \bar{Q} '_{pr, 1} / \bar{Q}' _{ext} ) ~\vec{v} / c  \right \}$ ~.

Since $\bar{Q} '_{pr, 1} / \bar{Q}' _{ext}$ $\ne$ 1 in general, also radial
term is present in $d \vec{p}_{out} / dt$, not only "forward direction", in
the inertial frame of the Sun.

As for the perfectly absorbing spherical particle:
the standard approach $\bar{Q} '_{pr, 1}$ $=$ $\bar{Q}' _{ext}$ $=$ 1
would yield
$d \vec{p}_{out} / d t$ $=$ ( $S ~A' / c$ ) $\vec{v} / c$ $=$
( $S ~A' / c$ ) $\left \{ ( v_{R} / c ) \vec{e}_{R} ~+~
( v_{T} / c ) \vec{e}_{T} \right \}$, while the correct result is
$\bar{Q} '_{pr, 1}$ $=$ 1, $\bar{Q}' _{ext}$ $=$ 2, and,
$d \vec{p}_{out} / d t$ $=$ ( $S ~A' / c$ ) $\left \{ \vec{e}_{R} ~+~
( v_{T} / c ) \vec{e}_{T} \right \}$; the velocity vector $\vec{v}$ is
decomposed into radial component $v_{R} \vec{e}_{R}$ and transversal
component $v_{T} \vec{e}_{T}$, $\vec{e}_{R}$ $\equiv$ $\vec{e}_{1}$.

iv) \\
"The particle thereby preferentially radiates
in the forward direction in the inertial frame of the Sun.
This causes dust in bound orbits to spiral sunward."
(Lissauer and Murray 2007, p. 806)

As it is evident from Eq. (110), the special case
$\bar{Q} '_{pr, 1} / \bar{Q}' _{ext}$ $=$ 1 yields
$( d \vec{v} / d t )_{out}$ $=$ 0 and the reradiated energy does not cause
dust to spiral sunward. Dealing with more realistic case
$\bar{Q} '_{pr, 1} / \bar{Q}' _{ext}$ $\ne$ 1, one obtains: \\
a) if $\bar{Q} '_{pr, 1} / \bar{Q}' _{ext}$ $<$ 1, then the re-radiated
light yields $(d \vec{v} / d t)_{out}$ $\propto$ $+$ $\vec{v}$, \\
b) if $\bar{Q} '_{pr, 1} / \bar{Q}' _{ext}$ $>$ 1, then the re-radiated
light yields $(d \vec{v} / d t)_{out}$ $\propto$ $-$ $\vec{v}$. \\
See also Sec. 7.4.1. Moreover, if optical properties of the dust change,
then the P-R effect can cause spiralling outward from the Sun.

v) \\
"The net force on a rapidly rotating dust grains is given by ... (Eq. 50-LM)"
(Lissauer and Murray 2007, p. 806)

In reality, Eq. (50-LM) holds for nonrotating spherical particle, as it
is evident from derivations presented in our paper (see also Sec. 7.3.3).
If we would like to consider rapidly rotating arbitrarily shaped dust particle,
we have to consider general equation of motion (see Eq. 39) and the
corresponding averaging over rotational motion has to be done (see, e. g.,
Kla\v{c}ka and Kocifaj 2001). Krauss and Wurm (2004) have given an
experimental evidence that arbitrarily
shaped and rapidly rotating dust grain moves in a different way than it
corresponds to the Poynting-Robertson effect.

vi) \\
Equation (50-LM) contains general factor
$Q_{pr}$ ($\equiv \bar{Q}' _{pr, 1}$). However, this is in contradiction
with the statement "A small particle in orbit
around the Sun absorbs solar radiation and reradiates the energy isotropically
in its own frame." (Lissauer and Murray 2007, p. 806).
The isotropically reradiated energy within geometrical optics approximation
(diffraction is not considered in this formulation) corresponds to the condition
$\bar{Q} '_{pr, 1} / \bar{Q}' _{ext}$ $=$ 1/2 (see Eq. 104 in Sec. 7.3.3),
in reality. Thus, if the authors state that
the essence of the Poynting-Robertson effect is "particle reradiates the
energy isotropically in its own frame", then they have to use
$Q_{pr}$ $=$ 1.

It is also important to stress that the process "particle reradiates the energy
isotropically in its own frame" (Lissauer and Murray 2007, p. 806) is only
an approximation within geometrical optics and that, in reality, the fundamental
condition for the case is given by the equation $\vec{p}_{o} '$ $=$ (1/2)
$\vec{p}_{i} '$ and not $\vec{p}_{o} '$ $=$ 0. The last equation considers
the effect of diffraction (calculation of mass of the outgoing radiation
within relativity approach requires consideration of the effect of diffraction),
while the phrase "particle reradiates the energy isotropically in its own frame"
does not consider the effect of diffraction.

vii) \\
"The first term in Eq. (50-LM) is that due to radiation pressure and the
second and the third terms (those involving the velocity of the particle)
represent the Poynting-Robertson drag."
(Lissauer and Murray 2007, p. 806)

The effect of electromagnetic radiation on spherical dust particle is given
by Eq. (70). It corresponds to radiation pressure. Eq. (70) contains all the
terms present in Eq. (50-LM). Thus, Eq. (50-LM) corresponds to the
radiation pressure -- not only its first term, as the authors state.
Physics does not allow to divide Eq. (70) into several parts which can be
treated as separate physical phenomena.

viii)
"From this discussion, it is clear that small-sized dust grains in the
interplanetary medium are removed: (sub)-micronsized grains are blown out
of the solar system, whereas larger particles spiral inward toward the Sun."
(Lissauer and Murray 2007, p. 806)

The conclusion made by the authors does not follow from the discussion
presented by the authors. Even if we take into account all the corrections
presented above and that the correct and fundamental result is given by
Eq. (50-LM). The authors' conclusion may be violated if one takes into
account that optical properties of the particles may change.

\subsubsection{Aberration (of light)}
It is often stated that the aberration of the incoming light is responsible
for the term $-$ $\vec{v} / c$ in the Poynting-Robertson effect
(see, e. g., Dohnanyi 1978, p. 562; Leinert and Gr\H{u}n 1990, p. 226;
Gr\H{u}n 2007, p. 632).
Is it true? If the incoming radiation would be responsible
for the term $-$ $\vec{v} / c$ in the P-R effect ("radiation falls
preferentially on the leading edge of the orbiting particle and acts as a drag
force" -- Festou {\it et al.} 2004, p. 729), then this term should
be present in $\vec{p}_{i}$ $\equiv$ $d \vec{p}_{in} / d \tau$.
However, Eqs. (42) and (75) show that $d \vec{p}_{in} / d \tau$
is proportional to $\vec{b}_{1}$ $=$ $\vec{e}_{1} / w_{1}$. Thus, the term
$-$ $\vec{v} / c$ does not come from the incoming radiation. In reality,
the term $-$ $\vec{v} / c$ is present in $d \vec{p}_{out} / d \tau$
(Eq. 76). But it's presence is not due to the aberration of light. It's
existence is caused by conservation of particle's mass, when considering both
the incoming and the outgoing radiation. Really, if we would require, e. g.,
$d E' / d \tau$ $=$ ($w_{1}^{2}$ $S$ $\bar{Q}'_{pr}$ $A'$ $/$ $c$)
$\alpha'$ and $d \vec{p}' / d \tau$ $=$ ($w_{1}^{2}$ $S$ $\bar{Q}'_{pr}$
$A'$ $/$ $c$) $\vec{e}_{1} '$, then Eq. (14) would yield
$d \vec{p} / d \tau$ $=$ ($w_{1}^{2}$ $S$ $\bar{Q}'_{pr}$ $A'$ $/$ $c$)
[ $\vec{b}_{1}$ $+$ ( $\alpha'$ $-$ 1 ) $\vec{u} / c$ ], where
$\vec{b}_{1}$ $=$ $\vec{e}_{1} / w_{1}$, $\vec{u}$ $=$ $\gamma$ $\vec{v}$,
$w_{1}$ $=$ $\gamma$ ( 1 $-$ $\vec{v}$ $\cdot$ $\vec{e}_{1}$ $/$ $c$ ), and,
$\alpha'$ is a constant. The term ( $\alpha'$ $-$ 1 ) $\gamma \vec{v} / c$
does not correspond to the aberration of light. As a similar example, we can
mention Eq. (40).

\subsection{Analogy between mechanics and electromagnetism?}
We have discussed the effect of electromagnetic radiation (photons) on
perfectly absorbing and totally reflecting spherical and planar particles.
We have stressed that even within the geometrical optics approximation
the effect of diffraction is important, in the process of interaction between
the particle and the electromagnetic radiation. As a consequence, the
relation between the outgoing and incoming momenta per unit time (Eq. 73)
reduces to $\vec{p}_{o}'$ $=$ 0.5 $\vec{p}_{i}'$
for the above mentioned spherical particles (Eq. 104). The numerical factor
0.5 considers also the effect of diffraction which is conventionally
neglected (e. g., Lissauer and Murray 2007).

Thus, we know that "for large spheres we may separate the effects of
geometrical optics (reflection and refraction) and the effect of diffraction"
(van de Hulst 1981, p. 225). But even in the case of geometrical optics approach,
the simultaneous action of these effects cannot be neglected, if one wants to be
consistent with the relativity theory. If we take into account a plane
particle (arbitrarily shaped particle, in general),
all these effects are also important.

The effect of reflection
of the electromagnetic radiation on a planar object is often explained
in an analogy to mechanical process. As an example, we can mention
two well-known textbooks on fundamental physics. Halliday {\it et al.}
(2008, p. 900) write about the comparison between the radiation effect
on perfectly absorbing and totally reflecting object: "Instead of being
absorbed, the radiation can be {\bf reflected} by the object: that is,
the radiation can be sent off in a new direction as if it bounced off
the object. If the radiation is entirely reflected back along its original
path, the magnitude of the momentum change of the object is twice that
given above ... (the case for total absorption)". "In the same way, an object
undergoes twice as much momentum change when a perfectly elastic tennis ball
is bounced from it as when it is struck by a perfectly inelastic ball
(a lump of wet putty, say) of the same mass and velocity."
Similarly, Jewett and  Serway (2008, p. 963) write: "Electromagnetic
waves transport linear momentum as well as energy. ... In this discussion,
let's assume the electromagnetic wave strikes the surface at normal
incidence ..." "Momentum transported to the perfectly absorbing surface
has a magnitude $p$." The authors continue (Jewett and  Serway 2008, p. 964):
"If the surface is a perfect reflector (such as a mirror) and incidence is
normal, the momentum transported to the surface ... is twice that given by
(the magnitude $p$). That is, the momentum transferred to the surface by the
incoming light is $p$ and that transferred to the surface by the
reflected light also is $p$".  In other words, the momentum transferred to the
surface by the incoming light equals the momentum transferred by the
reflected light. If the mechanical process would be a good analogue to the
electromagnetic process, then the relation $\vec{p}_{o}'$ $=$ ( 1 $-$
$\bar{C}_{pr, 1} '$ $/$ $\bar{C}_{ext} '$) $\vec{p}_{i}'$ would reduce to
$\vec{p}_{o}'$ $=$ $-$ $\vec{p}_{i}'$, for the perfect
(both mechanical and electromagnetic) reflection. But the reality is different.
The mechanical process is characterized by the values
$\bar{C}_{pr, 1} '$ $=$ 2 $A'$, $\bar{C}_{ext} '$ $=$ $A'$,
while the electromagnetic process, including the effect of diffraction,
yields cross sections
$\bar{C}_{pr, 1} '$ $=$ 2 $A'$, $\bar{C}_{ext} '$ $=$ 2 $A'$, where $A'$
is geometrical cross section of the object.

Summarization of the incident and outgoing momenta per unit time
can be done in the following way ($A'$ is geometrical cross section of the
planar surface, $u'$ and $c$ are the speeds of the incident mechanical particle
and photon with respect to the surface, and, $S'$ is the rate of flow of
energy, i. e., the rate at which the energy flows through a unit surface area
perpendicular to the direction of the particle or photon propagation),
for normal incident energy: \\
i) {\it perfectly absorbing planar surface}: \\
$\bullet$ {\bf mechanics:}
$\vec{p}_{i}'$ $=$ ( $S' ~A' / c$ ) ( $u'$ $/$ $c$ ) $\vec{e}_{1}'$,
$\vec{p}_{o}'$ $=$ 0 \\
$\bullet$ {\bf electromagnetism:}
$\vec{p}_{i}'$ $=$ 2 ( $S' ~A' / c$ ) $\vec{e}_{1}'$,
$\vec{p}_{o}'$ $=$ ( $S' ~A' / c$ ) $\vec{e}_{1}'$ \\
$[$ 1 $\times$ ( $S' ~A' / c$ ) $\vec{e}_{1}'$ comes from the effect of
diffraction, simultaneously in $\vec{p}_{i}'$ and $\vec{p}_{o}'$ $]$ \\
ii) {\it perfectly reflecting planar surface}: \\
$\bullet$ {\bf mechanics:}
$\vec{p}_{i}'$ $=$ ( $S' ~A' / c$ ) ( $u'$ $/$ $c$ ) $\vec{e}_{1}'$,
$\vec{p}_{o}'$ $=$ $-$ ( $S' ~A' / c$ ) ( $u'$ $/$ $c$ ) $\vec{e}_{1}'$ \\
$\bullet$ {\bf electromagnetism:}
$\vec{p}_{i}'$ $=$ 2 ( $S' ~A' / c$ ) $\vec{e}_{1}'$,
$\vec{p}_{o}'$ $=$ 0 \\
$[$ as for the $\vec{p}_{o}'$: the effect of diffraction
( $S' ~A' / c$ ) $\vec{e}_{1}'$ is compensated by
the effect of reflection $-$ ( $S' ~A' / c$ ) $\vec{e}_{1}'$ $]$ . \\
Finally, we will present a short explanation to the results of mechanics.
Four-momentum of the incident (classical) mechanical particle is
$p'^{\mu}_{MP}$ $=$ ( $E'_{MP} / c$, $\vec{p}'_{MP}$ ). Four-momentum per unit
time is $p'^{\mu}_{i}$ $=$ ( $E'_{i} / c$, $\vec{p}'_{i}$ ) $=$ $n'$ $u'$ $A'$
$p'^{\mu}_{MP}$ $=$ ( $n'$ $u'$ $A'$ $E'_{MP} / c$ ) $\times$
( 1, $\vec{u}' / c$ ), since $\vec{p}'_{MP}$ $=$ $E'_{MP}$ $\vec{u}'$ $/$
$c^{2}$, $\vec{u}'$ $=$ $u'$ $\vec{e}_{1}'$. Defining $S'$ $=$ $n'$ $u'$
$E'_{MP}$, we obtain $p'^{\mu}_{i}$ $=$ ( $S'$ $A'$ $/$ $c$ ) $\times$
( 1, $u'$ $\vec{e}_{1}'$ $/$ $c$ ).

\section{On the equivalence principle of mass and energy}
One of the well-known results of the special relativity theory is {\it the
equivalence of mass and energy} (more properly: the relation between mass
and energy). This "universal equivalence principle of mass
and energy" is often formulated in the following way
(or a little different, but equivalent form): "there does not exist a mass
without energy (and vice versa) and every change in energy is connected with
a corresponding change of inertial mass" (Schr\"{o}der 1990, p. 114).
It is stated that it holds also for a photon (light) and an inertial mass
of the photon equals $m_{photon}$ $=$ $E/c^{2}$, where $E$ is energy of the
photon. A simple "proof" can be found in Kittel {\it et al.} (1962; Sec. 12.6,
Eqs. 58-59, p. 400): $p = m_{photon} c$ and $E/c = p$, which yields
$E = m_{photon} c^{2}$, $p$ is momentum of the photon.
(Serway {\it et al.} 2005, p. 95: "The photon has zero mass, but its effective
inertial mass $m_{i}$ may reasonably be taken to be the mass equivalent of the
photon energy $E$, or $m_{i}$ $=$ $E / c^{2}$ $=$ $h f / c^{2}$. The same result
is obtained if we divide the photon momentum by the photon speed $c$:
$m_{i}$ $=$ $p / c$ $=$ $h f / c^{2}$." Analogously, Gasiorowicz 1974, p. 466
states: "A photon of energy $E$ has gravitational mass $E / c^{2}$.")
The principle is usually proved by the thought experiment, coming back to
Einstein (see, e. g., Einstein 1906, p. 633; Beiser 1969, Sec. 2.6;
Bernstein {\it et al.} 2000, pp. 81-82; Chow 2008, pp. 272-273;
compare also Okun 1989b, p. 34). The idea is following:
"Classical electromagnetism tells us that light carries both momentum
and energy. Because of this momentum, light exerts a pressure. We can think
of the radiation as being composed of massless particles -- any object
that moves with the speed of light must have no mass -- carrying momentum.
.... A box on a frictionless surface has a radiation emitter on one end
and an absorbing surface on the other. When a burst of radiation is
emitted, the box recoils, stopping only when the radiation is absorbed at
the other end. The center of mass will not move if there is mass-energy
equivalence." (Bernstein {\it et al.} 2000, p. 81). As a result, the relation
for a beam of parallelly spreading photons is presented: $E$ $=$ $m$ $c^{2}$,
where $E$ is energy of the photons (electromagnetic radiation), $c$ is the
speed of light and $m$ is the mass corresponding to the energy $E$
(the result of the thought experiment is: "The energy of the radiation is
equivalent to a mass according to the Einstein formula: $E$ $=$ $m$ $c^{2}$."
Bernstein {\it et al.} 2000, p. 82).

However, the following question arises: What is the physical sense of the
mass for a beam of masless photons? According to Schr\"{o}der (1990, p. 106):
"The inertia of the particle is described by an invariant quantity, the mass
$m$ of the particle, more properly called its rest mass". On the other hand,
the author writes: "According to the universal equivalence principle
of mass and energy, there does not exist a mass without energy (or vice versa)
and every change in energy is connected with a corresponding change of
inertial mass." (Schr\"{o}der 1990, p. 114). Thus, an inconsistency emerges:
the inertia of a beam of massless particles, photons, is described by an
invariant quantity, the mass of the beam $m_{beam}$ $=$ 0, and, the beam
possesses and energy $E_{beam}$ which should be equivalent to a
mass according to the Einstein formula $m_{beam}$ $=$ $E_{beam}$ $/$ $c^{2}$.

\subsection{Thought experiment}
In trying to better understand the situation, we wil present detailed
calculations for a thought experiment. The thought experiment will be
similar to that discussed above, but instead of a box of mass $M$ we will
consider two surfaces of masses $M/2$ on a frictionless surface. The
surfaces are penpendicular to the frictionless surface. They are at rest
in an inertial frame of reference $K$, at the beginning, with initial positions
$x_{L}$ $=$ $- L / 2$ and $x_{R}$ $=$ $+ L / 2$. The right surface emits
a particle with energy $E$ (measured in the reference frame $K$) towards
the left surface, i. e. direction and orientation of the motion of the particle
is characterized by the unit vector $-$ $\hat{\vec{x}}$. Let the emission of
the energy $E$ corresponds to a mass decrease $m_{R}$ of the right surface.
The absorption of the energy $E$ produces a mass increase $m_{L}$ of the
left surface.

\subsubsection{Emission of a massless particle -- particle with zero mass}
After emission of a massless particle (photon) by the right surface and its
absorption by the left surface, the center of inertia of the system
does not change (see, e. g., Eq. 14.6 in Landau and Lifshitz 2005):
\begin{equation}\label{114}
E_{L} ~\vec{r}_{L} ~+~ E_{R} ~\vec{r}_{R} = 0 ~.
\end{equation}
We can rewrite Eq. (114) in detail:
\begin{equation}\label{115}
\gamma_{L} \left ( \frac{M}{2} ~+~ m_{L} \right ) ~\left ( \frac{L}{2} ~+~
v_{L} ~t \right ) =
\gamma_{R} \left ( \frac{M}{2} ~-~ m_{R} \right ) ~\left ( \frac{L}{2} ~+~
v_{R} ~ \frac{L}{c} ~+~ v_{R} ~t \right ) ~,
\end{equation}
where $t =$ 0 holds for the moment of the photon absorption by the left
surface, $\gamma_{L}$ $=$ $1/ \sqrt{1 - (v_{L} / c)^{2}}$,
$\gamma_{R}$ $=$ $1/ \sqrt{1 - (v_{R} / c)^{2}}$; the multiplicative factors
$c^{2}$ present in energies
$E_{L}$ $=$ $\gamma_{L} ( M/2 ~+~ m_{L}) ~c^{2}$ and
$E_{R}$ $=$ $\gamma_{R} ( M/2 ~-~ m_{R}) ~c^{2}$ are omitted.
Eq. (109) yields
\begin{equation}\label{116}
\gamma_{L} \left ( \frac{M}{2} ~+~ m_{L} \right ) =
\gamma_{R} \left ( \frac{M}{2} ~-~ m_{R} \right ) ~ \left ( 1~+~
2 ~ \frac{v_{R}}{c}  \right )
\end{equation}
for $t =$ 0. Since the center of inertia does not change also for $t >$ 0,
Eq. (115) yields
\begin{equation}\label{117}
\gamma_{L} \left ( \frac{M}{2} ~+~ m_{L} \right ) ~ v_{L}  =
\gamma_{R} \left ( \frac{M}{2} ~-~ m_{R} \right ) ~ v_{R} ~.
\end{equation}

Eq. (117) corresponds to the laws of conservation of momentum for the
moments of emission and absorption:
\begin{equation}\label{118}
\gamma_{R} \left ( \frac{M}{2} ~-~ m_{R} \right ) ~ v_{R}  = \frac{E}{c} ~,
\end{equation}
\begin{equation}\label{119}
\gamma_{L} \left ( \frac{M}{2} ~+~ m_{L} \right ) ~ v_{L}  = \frac{E}{c} ~.
\end{equation}

The laws of conservation of energy for the emission and the absorption
are of the following forms in the inertial frame of reference $K$:
\begin{equation}\label{120}
\gamma_{R} \left ( \frac{M}{2} ~-~ m_{R} \right ) ~c^{2} ~+~ E =
\frac{M}{2} ~c^{2} ~,
\end{equation}
\begin{equation}\label{121}
\gamma_{L} \left ( \frac{M}{2} ~+~ m_{L} \right ) ~c^{2} =
\frac{M}{2} ~c^{2} ~+~ E ~.
\end{equation}

Eqs. (118) and (120) yield
\begin{equation}\label{122}
m_{R} = \frac{M}{2} ~ \left \{ 1 ~-~ \sqrt{1 ~-~ 2~ \frac{E / c^{2}}{M / 2 }}
	\right \} ~,
\end{equation}
or,
\begin{equation}\label{123}
E = \left ( 1 ~-~ \frac{m_{R}}{M} \right ) ~m_{R} ~ c^{2} ~.
\end{equation}
During the process of the emission of the photon(s), the mass $m_{R}$
(more correctly: the energy $m_{R}$ $c^{2}$) has been changed into the
energy $E$ and kinetic energy of the surface
of mass $M/2 - m_{R}$ [kinetic energy $=$ ($\gamma_{R}$ $-$ 1) ($M/2 - m_{R}$)
$c^{2}$]; $m_{R}$ $\approx$ [ 1 + ( $E/c^{2}$) $/$ $M$] $E/c^{2}$. This is in
agreement with Eq. (120) and this is the physical interpretation
of the Eqs. (118), (120) and (122)-(123). Mass of the photon(s) with the
energy $E$ is zero.

Eqs. (119) and (121) yield
\begin{equation}\label{124}
m_{L} = \frac{M}{2} ~ \left \{ \sqrt{1 ~+~ 2~ \frac{E / c^{2}}{M / 2 }}
	~-~ 1 \right \} ~,
\end{equation}
or,
\begin{equation}\label{125}
E = \left ( 1 ~+~ \frac{m_{L}}{M} \right ) ~m_{L} ~ c^{2} ~.
\end{equation}
During the process of the absorption of the photon(s), the energy $E$ has been
changed into the mass $m_{L}$ (more correctly: rest energy $m_{L}$ $c^{2}$)
and motion of the surface of mass $M/2 + m_{L}$
[kinetic energy $=$ ($\gamma_{L}$ $-$ 1) ($M/2 + m_{L}$) $c^{2}$];
$m_{L}$ $\approx$ [ 1 $-$ ( $E/c^{2}$ ) $/$ $M$ ] $E/c^{2}$. This is in
agreement with Eq. (121) and this is the physical interpretation
of Eqs. (119), (121) and (124)-(125).

\subsubsection{Summary}
We can make the following statements: \\
i) Photon(s) (light) with the energy $E$ transferred mass $m_{L}$. \\
i) Mass of the photon(s) (light) is zero. \\
Proof of the statements: \\
i) Initial mass of the left surface was $M/2$, it's mass after absorbing the
photon(s) is $\sqrt{(energy / c^{2})^{2} - (momentum / c)^{2}}$ $=$
$\sqrt{[\gamma_{L} (M/2 + m_{L})]^{2} - [\gamma_{L} (M/2 + m_{L}) v_{L} / c]^{2}}$
$=$ $M/2 + m_{L}$. The mass $m_{L}$ is given by Eq. (117), or, approximately,
$m_{L}$ $\approx$ [ 1 $-$ ($E/c^{2}$) $/$ $M$] $E/c^{2}$. \\
ii) Mass of the photon(s) equals
$\sqrt{(energy / c^{2})^{2} - (momentum / c)^{2}}$ $=$
$\sqrt{(E / c^{2})^{2} - (\vec{p} / c)^{2}}$ $=$
$\sqrt{(E / c^{2})^{2} - [(E / c) ( - \hat{\vec{x}} ) / c]^{2}}$ $=$
$\sqrt{(E / c^{2})^{2} - (E / c^{2})^{2}}$ $=$ 0, where also the fact that
the photon(s) moves with the speed $c$ in the direction and orientation
$- \hat{\vec{x}}$, was used.

The statements are consistent with the statements presented in
Taylor and Wheeler (1992, p. 228): "Energy without mass: photon",
"photon moves with zero mass".

\subsubsection{Emission of a particle with non-zero mass}
After emission of a particle with non-zero mass by the right surface and its
absorption by the left surface, the center of inertia of the system
does not change, see Eq. (114).
We can rewrite Eq. (114) in detail:
\begin{equation}\label{126}
\gamma_{L} \left ( \frac{M}{2} ~+~ m_{L} \right ) ~\left ( \frac{L}{2} ~+~
v_{L} ~t \right ) =
\gamma_{R} \left ( \frac{M}{2} ~-~ m_{R} \right ) ~\left ( \frac{L}{2} ~+~
v_{R} ~ \frac{L}{u} ~+~ v_{R} ~t \right ) ~,
\end{equation}
where $t =$ 0 holds for the moment of the absorption of the particle
by the left surface, $\gamma_{L}$ $=$ $1/ \sqrt{1 - (v_{L} / c)^{2}}$,
$\gamma_{R}$ $=$ $1/ \sqrt{1 - (v_{R} / c)^{2}}$, $u$ is the speed of the
particle in the frame of reference $K$;
the multiplicative factors $c^{2}$ present in energies
$E_{L}$ $=$ $\gamma_{L} ( M/2 ~+~ m_{L}) ~c^{2}$ and
$E_{R}$ $=$ $\gamma_{R} ( M/2 ~-~ m_{R}) ~c^{2}$ are omitted.
Eq. (126) yields
\begin{equation}\label{127}
\gamma_{L} \left ( \frac{M}{2} ~+~ m_{L} \right ) =
\gamma_{R} \left ( \frac{M}{2} ~-~ m_{R} \right ) ~ \left ( 1~+~
2 ~ \frac{v_{R}}{u}  \right )
\end{equation}
for $t =$ 0. Since the center of inertia does not change also for $t >$ 0,
Eq. (126) yields
\begin{equation}\label{128}
\gamma_{L} \left ( \frac{M}{2} ~+~ m_{L} \right ) ~ v_{L}  =
\gamma_{R} \left ( \frac{M}{2} ~-~ m_{R} \right ) ~ v_{R} ~.
\end{equation}

Eq. (128) corresponds to the laws of conservation of momentum for the
moments of emission and absorption:
\begin{equation}\label{129}
\gamma_{R} \left ( \frac{M}{2} ~-~ m_{R} \right ) ~ v_{R}  = \gamma_{u} ~m~u ~,
\end{equation}
\begin{equation}\label{130}
\gamma_{L} \left ( \frac{M}{2} ~+~ m_{L} \right ) ~ v_{L}  = \gamma_{u} ~m~u  ~,
\end{equation}
where $m$ is mass of the particle and $\gamma_{u}$ $=$
$1 / \sqrt{1 - ( u / c )^{2}}$.

The laws of conservation of energy for the emission and the absorption
are of the following forms in the inertial frame of reference $K$:
\begin{equation}\label{131}
\gamma_{R} \left ( \frac{M}{2} ~-~ m_{R} \right ) ~c^{2} ~+~
    \gamma_{u} ~m ~c^{2} = \frac{M}{2} ~c^{2} ~,
\end{equation}
\begin{equation}\label{132}
\gamma_{L} \left ( \frac{M}{2} ~+~ m_{L} \right ) ~c^{2} =
\frac{M}{2} ~c^{2} ~+~ \gamma_{u} ~m ~c^{2} ~.
\end{equation}

Eqs. (129) and (131) yield
\begin{eqnarray}\label{133}
m_{R} &=& \frac{M}{2} ~ \left \{ 1 ~-~ \sqrt{1 ~-~ 2~ \frac{E / c^{2}}{M / 2}
	~+~ \left ( \frac{m}{M / 2} \right )^{2}} \right \} ~,
\nonumber \\
E &=& \gamma_{u} ~ m ~c^{2} ~,
\end{eqnarray}
or,
\begin{eqnarray}\label{134}
E &=& \left ( 1 ~-~ \frac{m_{R}}{M} \right ) ~m_{R} ~ c^{2} ~+~
      \frac{m}{M} ~ m ~c^{2}
\nonumber \\
  &=&  m_{R} ~ c^{2} ~-~
       \frac{m_{R}^{2} ~-~ m^{2}}{M} ~ c^{2}  ~,  ~~~ m <  m_{R} ~.
\end{eqnarray}
During the process of the emission of the massive particle, the energy $m_{R}$
$c^{2}$ has been changed into the energy $E$ and kinetic energy of the surface
of mass $M/2 - m_{R}$ [kinetic energy $=$ ($\gamma_{R}$ $-$ 1) ($M/2 - m_{R}$)
$c^{2}$]. This is in agreement with Eq. (131) and this is the physical
interpretation of Eqs. (129), (131) and (133)-(134). Mass of the massive
particle with the energy $E$ is $m$.

Eqs. (130) and (132) yield
\begin{eqnarray}\label{135}
m_{L} &=& \frac{M}{2} ~ \left \{ \sqrt{1 ~+~ 2~ \frac{E / c^{2}}{M / 2}
	~+~ \left ( \frac{m}{M / 2} \right )^{2}} ~-~ 1 \right \} ~,
\nonumber \\
E &=& \gamma_{u} ~ m ~c^{2} ~,
\end{eqnarray}
or,
\begin{eqnarray}\label{136}
E &=& \left ( 1 ~+~ \frac{m_{L}}{M} \right ) ~m_{L} ~ c^{2} ~-~
    \frac{m}{M} ~ m ~c^{2}
\nonumber \\
  &=&  m_{L} ~ c^{2} ~+~
       \frac{m_{L}^{2} ~-~ m^{2}}{M} ~ c^{2} ~, ~~~ m < m_{L} ~.
\end{eqnarray}
During the process of the absorption of the massive particle, the energy $E$ has
been changed into the mass $m_{L}$ (more correctly: rest energy $m_{L}$ $c^{2}$)
and motion of the surface of mass $M/2 + m_{L}$
[kinetic energy $=$ ($\gamma_{L}$ $-$ 1) ($M/2 + m_{L}$) $c^{2}$]. This is in
agreement with Eq. (130) and this is the physical interpretation
of the Eqs. (130), (132) and (135)-(136).

\subsection{Summary}
The important statements, as for massive particles, can be summarized as
follows: "The theory of relativity leads to the important conclusion that the
energy of a particle (with $m \ne$ 0) at rest is $m c^{2}$. The relations
$E_{0} = m c^{2}$, $E = m c^{2} / \sqrt{1 - ( u / c )^{2}}$ for a particle
at rest and a particle in motion, respectively, are the famous Einstein
formulae. They express the equivalence of mass and energy." (Schr\"{o}der 1990,
p. 113). As for the massless particle, it moves with the speed $c$ and, thus,
it cannot be at rest with respect to any frame of reference: rest energy
of the particle is $E_{0} =$ 0. The inertia of the particle is described by an
invariant quantity, the mass of the particle. The mass of a photon is zero.

The term "inertial mass of a photon" is of no physical sense: the speed
of the photon is constant and the photon cannot be accelerated or
decelerated, in a vacuum. Similarily, the term "gravitational mass of the
photon" is of no physical sense -- no (thought) experiment exists which would
prove the correctness of the gravitational character of the photon. (See
also Ugarov 1969, p. 248).

Finally, we will reproduce several sentences on the meaning of $E_{0}$ $=$
$m$ $c^{2}$ from Sachs (2007, p. 84): "It has been said by many physicists
and philosophers that the formula $E_{0}$ $=$ $m$ $c^{2}$ [we have added
the subscript $0$] means that `mass is equivalent to energy'. This is
philosophically false. It is not what Einstein said when he derived
this relation. What he said was that `the inertial mass of matter is
{\it a measure} of its energy content'. In physics, as Newton originally
postulated, the inertial mass of matter is, {\it by definition}, a measure
of its resistance to a change of its state of rest or constant motion.
The energy of matter, on the other hand, is {\it by definition}, the
capability of this matter to do work. Thus, mass and energy are totally
different concepts! What should be said, instead of saying that mass
is equivalent to energy, is that {\it mass} (the inertia of matter) is
a {\it measure} of the capability of this matter to do work (its intrinsic
{\it energy})."

Results obtained in Secs. 4-7 are consistent with the previous statements
(see also Eq. 105 with comment).

\subsection{Comparison of the results}
We want to compare the results obtained in Sec. 8.1 with the results of
Secs. 2-7.

\subsubsection{Relativity and optics from Secs. 2-7}
Let us consider an incoming electromagnetic radiation which interacts
with arbitrarily shaped particle. The incoming radiation is characterized by
its energy and momentum. The relevant relations are given by Eqs. (42), (50),
(60)-(63), or, by Eqs. (85)-(86) for spherical particles. In order to be the
results easily comparable to the considerations presented in Sec. 8.1.1, we will
discuss a totally absorbing planar surface of geometrical cross section $A'$
and normal incident radiation, within geometrical optics approximation. On the
basis of Secs. 7.3.3 (Eqs. 104-105 and the text above and below them) and 7.5,
we can immediately write
\begin{eqnarray}\label{137}
C'_{ext} &=& 2 ~A' ~,
\nonumber \\
C'_{sca} &=& A' ~,
\nonumber \\
E'_{i} &=& S' ~ C'_{ext} = 2~ S'~ A' ~,
\nonumber \\
\vec{p}'_{i} &=& \frac{E'_{i}}{c} ~\vec{e}_{1} ' ~,
\nonumber \\
dM_{in} &=& 0 ~,
\end{eqnarray}
where $S'$ is the flux of radiation energy and four-momentum per unit time is
$p_{i}^{' \mu}$ $=$ ( $E'_{i} / c$, $\vec{p}'_{i}$ ); $C'_{ext}$ and $C'_{sca}$
are cross sections for extinction and scattering.

If the absorption and thermal emission are present, then the outgoing
radiation is characterized by the following relations
\begin{eqnarray}\label{138}
E'_{o} &=& E'_{i}
\nonumber \\
C'_{pr, ~1} &=& A' ~,
\nonumber \\
\vec{p}'_{o} &=& \left ( 1 ~-~ \frac{C'_{pr, ~1}}{C'_{ext}} \right ) ~
		 \vec{p}'_{i} =  \frac{1}{2} ~ \vec{p}'_{i} ~,
\nonumber \\
dM_{out} &=& \sqrt{3} ~ \frac{S'~ A'}{c^{2}} ~ d \tau ~,
\end{eqnarray}
if the results from Eq. (137) are used.

Now, let the thermal emission does not exist (thought experiment).
Then
\begin{eqnarray}\label{139}
E'_{o} &=& 0 ~,
\nonumber \\
\vec{p}'_{o} &=& 0 ~,
\nonumber \\
C'_{pr, ~1} &=& 2 ~A' ~,
\nonumber \\
dM_{out} &=& 0 ~,
\end{eqnarray}
since $p_{o}^{' \mu}$ $p_{o~\mu} '$ $\ge$ 0,
$p_{o}^{' \mu}$ $=$ ( $E'_{o} / c$, $\vec{p}'_{o}$ ),
$\vec{p}'_{o}$ $=$ ( 1 $-$ $C'_{pr, ~1}$ $/$ $C'_{ext}$) $\vec{p}'_{i}$.
Eqs. (50) and (61)-(63) yield
\begin{eqnarray}\label{140}
dM_{in} &=& 0, ~
\nonumber \\
\left ( \frac{d m}{d \tau} \right ) _{in} &=&
		     \frac{w_{1}^{2} ~ S ~ C'_{ext}}{c^{2}} ~,
\nonumber \\
\frac{d E_{in}}{d \tau} &=&
\left ( \frac{d m}{d \tau} \right ) _{in} ~c^{2} ~ \frac{1}{w_{1}} ~,
\nonumber \\
\frac{d \vec{p}_{in}}{d \tau} &=& \left ( \frac{d m}{d \tau} \right )_{in} ~
				c ~\frac{\vec{e}_{1}}{w_{1}} ~,
\nonumber \\
w_{1} &=& \gamma_{v} ~( 1 ~-~ v / c ) ~,
\nonumber \\
C_{ext} ' &=& 2~ A' ~,
\end{eqnarray}
where also some results of Eq. (137) have been added.

\subsubsection{Relativity from Sec. 8.1.1}
Sec. 8.1.1 presents result for an increase of mass of a planar
absorbing surface if it's mass is $M/2$ and the increase of it's mass
is generated by an absorption of a photon of energy $E$ with respect to
the surface. In order to be able to compare the results of Sec. 8.1.1
with the results presented in Sec. 8.3.1, we have to generalize the
considerations of Sec. 8.1.1 for the case of a moving absorbing surface.

Let us consider a photon of energy $E$ in the laboratory frame (inertial
frame of reference $K$) in which the absorbing surface is moving with
the speed $v$, in the same direction and orientation as the photon is
travelling (also negative orientation may be considered: it is sufficient
to put $-$ $v$ into the formulae presented below, in this subsection).
We are interested in the increase of mass of the absorbing surface.
The law of conservation of momentum for the absorption
is of the following form in the inertial frame of reference $K$
(compare with Eq. 119):
\begin{equation}\label{141}
\gamma_{L} \left ( \frac{M}{2} ~+~ m_{L} \right ) ~ v_{L}  = \frac{E}{c}
	   ~+~ \gamma_{v} ~\frac{M}{2} ~ v ~.
\end{equation}
Similarily, the law of conservation of energy yields (compare with Eq. 121):
\begin{equation}\label{142}
\gamma_{L} \left ( \frac{M}{2} ~+~ m_{L} \right ) ~c^{2} = E ~+~
		 \gamma_{v} ~\frac{M}{2} ~c^{2} ~.
\end{equation}
Eqs. (141)-(142) yield
\begin{eqnarray}\label{143}
m_{L} &=& \frac{M}{2} ~ \left \{ \sqrt{1 ~+~ 2~ w_{1} ~\frac{E / c^{2}}{M / 2 }}
	~-~ 1 \right \} ~,
\nonumber \\
w_{1} &=& \gamma_{v} ~\left ( 1 ~-~ \frac{v}{c} \right ) ~.
\end{eqnarray}
During the process of the absorption of the photon(s), the energy $E$ has been
changed into the mass $m_{L}$ (more correctly: rest energy $m_{L}$ $c^{2}$)
and motion of it, and, part of the energy $E$ was changed into the change
of the kinetic energy of the surface of mass $M/2$. This can be written as:
$E$ $=$ $m_{L}$ $c^{2}$ $+$ ($\gamma_{L}$ $-$ 1) ($M/2 + m_{L}$) $c^{2}$ $-$
($\gamma_{v}$ $-$ 1) ($M/2$)  $c^{2}$, or,
$E$ $=$ $\gamma_{L}$ $m_{L}$ $c^{2}$ $+$
($\gamma_{L}$ $-$ $\gamma_{v}$) ($M/2$) $c^{2}$.

The limiting case $M \rightarrow \infty$ of Eq. (143) yields
\begin{eqnarray}\label{144}
\lim_{M \rightarrow \infty} m_{L} &=& w_{1} ~\frac{E}{c^{2}} ~,
\nonumber \\
w_{1} &=& \gamma_{v} ~\left ( 1 ~-~ \frac{v}{c} \right ) ~.
\end{eqnarray}

Eq. (144) can be written as
\begin{equation}\label{145}
\lim_{M \rightarrow \infty} m_{L} = \frac{E'}{c^{2}} ~,
\end{equation}
where $E'$ is energy of the photon in the frame of reference of the
absorbing surface.

\subsubsection{Relativity from Sec. 8.1.3}
Sec. 8.1.3 presents result for an increase of mass of a planar
absorbing surface if it's mass is $M/2$ and the increase of it's mass
is generated by an absorption of a massive particle of energy $E$ $=$
$\gamma_{u}$ $m$ $c^{2}$ with respect to the surface.
In order to be able to compare the results of Sec. 8.1.3 with
with the results presented in Sec. 8.3.2, we have to generalize the
considerations of Sec. 8.1.3 for the case of a moving absorbing surface.

Let us consider a massive particle of mass $m$ and energy $E$
in the laboratory frame (inertial frame of reference $K$)
in which the absorbing surface is moving with the speed $v$, in the same
direction and orientation as the massive particle is travelling, $u$ $>$ $v$
(also negative orientation may be considered: it is sufficient
to put $-$ $v$ into the formulae presented below, in this subsection:
since $u$ $>$ 0 also $u$ $>$ $-$ $v$).
We are interested in the increase of mass of the absorbing surface.
The law of conservation of momentum for the absorption gets
the following form in the inertial frame of reference $K$
(compare with Eq. 130):
\begin{equation}\label{146}
\gamma_{L} \left ( \frac{M}{2} ~+~ m_{L} \right ) ~ v_{L}  = \gamma_{u} ~m~ u
	   ~+~ \gamma_{v} ~\frac{M}{2} ~ v ~.
\end{equation}
Similarily, the law of conservation of energy yields (compare with Eq. 132):
\begin{equation}\label{147}
\gamma_{L} \left ( \frac{M}{2} ~+~ m_{L} \right ) ~c^{2} =
		 \gamma_{u} ~m ~c^{2} ~+~
		 \gamma_{v} ~\frac{M}{2} ~c^{2} ~.
\end{equation}
Eqs. (146)-(147) yield
\begin{eqnarray}\label{148}
m_{L} &=& \frac{M}{2} ~ \left \{ \sqrt{1 ~+~ 2 ~\gamma_{u} ~\gamma_{v} ~
	  \left ( 1 ~-~ \frac{u~v}{c^{2}} \right ) ~ \frac{m}{M / 2}
	  ~+~ \left ( \frac{m}{M / 2} \right )^{2}}
	  ~-~ 1 \right \} ~.
\end{eqnarray}
Eq. (148) may be rewritten also in the following forms:
\begin{eqnarray}\label{149}
m_{L} &=& \frac{M}{2} ~ \left \{ \sqrt{1 ~+~ 2~w_{u} ~
	  \frac{E / c^{2}}{M / 2}
	  ~+~ \left ( \frac{m}{M / 2} \right )^{2}}
	  ~-~ 1 \right \}
\nonumber \\
      &=& \frac{M}{2} ~ \left \{ \sqrt{1 ~+~ 2~
	  \gamma_{u'} ~ \frac{m}{M / 2} ~+~
	  \left ( \frac{m}{M / 2} \right )^{2}}
	  ~-~ 1 \right \} ~,
\nonumber \\
w_{u} &=& \gamma_{v} ~ \left ( 1 ~-~ \frac{u ~v}{c^{2}} \right )  ~,
\nonumber \\
E     &=& \gamma_{u} ~m ~c^{2} ~,
\nonumber \\
\gamma_{u'} &=& \gamma_{u} ~\gamma_{v} ~
		\left ( 1 ~-~ \frac{u~v}{c^{2}} \right ) ~,
\nonumber \\
u'    &=&  \frac{u ~-~ v}{1 ~-~ u~v ~/~c^{2}} ~.
\end{eqnarray}
The limiting case $M \rightarrow \infty$ of Eqs. (148)-(149) yields
\begin{eqnarray}\label{150}
\lim_{M \rightarrow \infty} m_{L} &=& \gamma_{u'} ~m
\nonumber \\
  &=& w_{u} ~\frac{E}{c^{2}} ~,
\nonumber \\
\gamma_{u'} &=& \gamma_{u} ~\gamma_{v} ~
		\left ( 1 ~-~ \frac{u~v}{c^{2}} \right ) ~,
\nonumber \\
w_{u} &=& \gamma_{v} ~\left ( 1 ~-~ \frac{u~v}{c^{2}} \right ) ~,
\nonumber \\
E     &=& \gamma_{u} ~m ~c^{2} ~.
\end{eqnarray}

Eq. (150) can be written as
\begin{equation}\label{151}
\lim_{M \rightarrow \infty} m_{L} = \frac{E'}{c^{2}} ~,
\end{equation}
where $E'$ is energy of the massive particle in the frame of reference of the
absorbing surface.

\subsubsection{Comparison of Secs. 8.3.1 and 8.3.2}
Let us consider a continuous beam of photons, each of which is bearing the
energy $E$ in the laboratory frame $K$. Let the absorbing surface is moving with
the speed $v$ in the laboratory frame, in the same direction and orientation
as the photons are travelling. The increase of mass of the absorbing surface,
per unit time, is
\begin{eqnarray}\label{152}
\left ( \frac{d m}{d \tau} \right ) _{in} &=& m_{L} ~\times~
		     \frac{S' ~ C'_{ext}}{E'} ~.
\end{eqnarray}
Now, we have to use Eqs. (19), (23) and (144). Eq. (152) leads to
\begin{eqnarray}\label{153}
\left ( \frac{d m}{d \tau} \right ) _{in} &=& w_{1} ~\frac{E}{c^{2}} ~ \times~
		     \frac{w_{1}^{2} ~S ~ C'_{ext}}{w_{1} ~E} =
\nonumber \\
&=&  \frac{w_{1}^{2} ~ S ~ C'_{ext}}{c^{2}} ~.
\end{eqnarray}
The result obtained in Eq. (153) is identical to the result presented in
Eq. (140) (see also Eq. 61). Thus, we have found a consistency between
considerations presented in Secs. 2-7 and Sec. 8.1.1. The term $w_{1}$
present in Eqs. (143)-(144) is due to the Doppler effect; the same
holds for the relations $dE_{in} / d\tau$ and $d\vec{p}_{in} / d\tau$
in Eq. (140).

\subsubsection{Comparison of Secs. 8.3.2 and 8.3.3}
Absorption of a photon bearing the energy $E$ in the laboratory frame $K$
leads to the increase of mass of the absorbing surface moving with the speed $v$
in the laboratory frame. The results are given by Eqs. (143)-(145).
Similarily, absorption of a massive particle (the particle with a non-zero mass)
bearing the energy $E$ in the laboratory frame $K$ also leads to the
increase of mass of the absorbing surface moving with the speed $v$
in the laboratory frame. The results are given by Eqs. (149)-(151).
Eqs. (145) and (151) are equivalent. Moreover, comparison of Eq. (144)
with Eq. (145) yields
\begin{equation}\label{154}
E' = \gamma_{v} ~ \left ( 1 ~-~ \frac{v}{c} \right ) ~ E ~,
\end{equation}
for the photon, and, comparison of Eq. (150) with Eq. (151) leads to
\begin{equation}\label{155}
E' = \gamma_{v} ~ \left ( 1 ~-~ \frac{u ~v}{c^{2}} \right ) ~ E ~,
\end{equation}
for the massive particle moving with the speed $u$ in the laboratory frame
of reference $K$. In both cases, $E$ is the energy of the incident photon /
massive particle in the laboratory frame of reference and $E'$ is the
energy of the photon / massive particle measured in the reference frame
of the absorbing surface. Eq. (154), which holds for the photon,
corresponds to the Doppler effect. Analogous equation for the massive
particle is Eq. (155) and it, formally, reduces to Eq. (154) in the
limiting case $u$ $\longrightarrow$ $c$.

According to Eq. (150), we have $\lim_{M \rightarrow \infty} m_{L}$ $=$
$\gamma_{u'}$ $m$, or, $\lim_{M \rightarrow \infty} m_{L}$ $=$
$w_{u}$ $E$ $/$ $c^{2}$. The quantity $m$ is the mass of the massive
particle and it is a relativistic invariant
($\sqrt{p_{\mu} ~p^{\mu}}$ $/$ $c$). The quantity $E$ $/$ $c^{2}$ is not
equal to $m$: $m$ $<$ $\lim_{M \rightarrow \infty} m_{L}$ $<$
$E$ $/$ $c^{2}$. Thus, it is of no sense to use $E$ $/$ $c^{2}$ as a mass
of the particle. As a consequence, also $E$ $/$ $c^{2}$ in Eq. (144) does
not represent a mass of a photon. These statements are consistent with
Sec. 8.2

\section{Cross sections for arbitrarily shaped particles}
We have already discussed, in Secs. 5, 6, 7.3 and 8, the relations between
four-momentum of the radiation and the mass of the radiation or the change
of the mass of the objects interacting with the radiation. Application
of these physical considerations to astrophysics was shown mainly in Sec. 7.4
for the case of dust grains with spherically symmetric mass distribution.
Discussion in Sec. 7.4 explained the physical access to the
Poynting-Robertson effect, which is standardly considered in evolution of
interplanetary dust grains. This section will discuss the relation
between four-momentum of the outgoing radiation and its mass for
arbitrarily shaped dust grains.

Let us consider that an arbitrarily shaped dust particle is irradiated
by the incoming radiation of a source of radiation. The relevant relations
for the four-momentum of the outgoing radiation and its mass are given by
equations presented in Sec. 5.2. In what follows, we want to obtain more
information on cross sections (for pressure components, extinction,
scattering) than can be obtained from observations and measurements.

As follows from Eqs. (53)-(54), the following inequality holds:
\begin{eqnarray}\label{156}
0 &\le& 2 \left (
	\frac{\bar{C}'_{pr, 1}}{\bar{C}'_{ext}} ~+~ X^{-1} ~ F'_{e, 1}
	\right ) ~-~ \sum_{j=1}^{3} \left (
	\frac{\bar{C}'_{pr, j}}{\bar{C}'_{ext}} ~+~ X^{-1} ~ F'_{e, j}
	\right )^{2}  ~,
\nonumber \\
X &\equiv& \frac{w_{1}^{2} ~ S ~ \bar{C}'_{ext}}{c} ~.
\end{eqnarray}
If we neglect the thermal emission force $\vec{F'_{e}}$, then
Eq. (156) reduces to
\begin{eqnarray}\label{157}
0 &\le& 2~ \frac{\bar{C}'_{pr, 1}}{\bar{C}'_{ext}}  ~-~
	\sum_{j=1}^{3} \left ( \frac{\bar{C}'_{pr, j}}{\bar{C}'_{ext}}
	\right )^{2}  ~.
\end{eqnarray}
Current situation is that we do not know simultaneous values of the cross
sections $\bar{C}'_{pr, j}$ ($j$ $=$ 1, 2, 3), $\bar{C}'_{ext}$,
$\bar{C}'_{sca}$ and $\bar{C}'_{abs}$ from observations or measurements
(Krauss and Wurm  2004). Moreover, if the size of the particle is larger
than about 5 micrometers, then it is not possible to make numerical
calculations for the above presented cross sections of the particle.
Thus, there do not exist methods which determine, simultaneously, the values
of cross sections
$\bar{C}'_{pr, j}$ ($j$ $=$ 1, 2, 3), $\bar{C}'_{ext}$, $\bar{C}'_{sca}$ and
$\bar{C}'_{abs}$, at present. However, we will show that some
information on the cross sections can be obtained on the basis of equations
presented in Secs. 2, 3.4 and 5.2, or, Eq. (157).

Let us suppose that measurements can yield ratios of the cross sections for
radiation pressure, i. e., the ratios $\bar{C}'_{pr, j}$ $/$
$\bar{C}'_{pr, 1}$ ($j$ $=$ 2, 3) are in disposal.
It is useful to rewrite Eq. (157) into the form
\begin{eqnarray}\label{158}
0 &<& \frac{\bar{C}'_{pr, 1}}{\bar{C}'_{ext}}  \le \frac{2}{1 ~+~
      \sum_{j=2}^{3} \left ( \bar{C}'_{pr, j} / \bar{C}'_{pr, 1}
      \right )^{2}} ~.
\end{eqnarray}
Eq. (158) reduces to Eq. (79) for the case
$\bar{C}'_{pr, 2}$ $=$ $\bar{C}'_{pr, 3}$ $=$ 0.
On the basis of Eqs. (9)-(10) and (38), we can write
$\bar{C}'_{pr, 1}$ $=$ $\bar{C}'_{ext}$ $-$
$\langle < \cos \theta'> ~ C'_{sca} \rangle$ (the outer symbol for the mean
denotes weighting over stellar spectrum and it corresponds
to the "bar" symbol used above the letters, standardly used in this paper).
Since $< \cos \theta'>$ $\le$ 1, we immediately obtain
\begin{eqnarray}\label{159}
\frac{\bar{C}'_{sca}}{\bar{C}'_{pr, 1}} &\ge&
\frac{\bar{C}'_{ext}}{\bar{C}'_{pr, 1}} ~-~ 1 ~.
\end{eqnarray}
The equation $\bar{C}'_{ext}$ $=$ $\bar{C}'_{abs}$ $+$ $\bar{C}'_{sca}$
can be rewritten into the form $\bar{C}'_{ext}$ $/$ $\bar{C}'_{pr, 1}$
$=$ $\bar{C}'_{abs}$ $/$ $\bar{C}'_{pr, 1}$ $+$
$\bar{C}'_{sca}$ $/$ $\bar{C}'_{pr, 1}$, which, together with Eq. (159)
yields
\begin{eqnarray}\label{160}
0 &\le& \frac{\bar{C}'_{abs}}{\bar{C}'_{pr, 1}} \le 1 ~.
\end{eqnarray}
Similarily, the relation $\bar{C}'_{ext}$ $/$ $\bar{C}'_{pr, 1}$
$=$ $\bar{C}'_{abs}$ $/$ $\bar{C}'_{pr, 1}$ $+$
$\bar{C}'_{sca}$ $/$ $\bar{C}'_{pr, 1}$
and Eq. (160) yield
\begin{eqnarray}\label{161}
0 &<& \frac{\bar{C}'_{sca}}{\bar{C}'_{pr, 1}} \le
\frac{\bar{C}'_{ext}}{\bar{C}'_{pr, 1}} ~.
\end{eqnarray}
Eqs. (158)-(161) offer some information on the cross sections of
extinction and scattering.

\section{Summary and conclusions}
The paper derives and presents relativistically covariant equation of motion
for dust particle under the action of electromagnetic radiation -- see
Eqs. (38) and (40). It yields, as special cases, the results obtained by
Einstein (1905) and Robertson (1937).
As for most frequent applications to systems in the universe
(e. g., meteoroids in the Solar System, dust particles in circumstellar disks),
equation of motion in the form of Eq. (39) is sufficient.

The general equation of motion reduces to the Poynting-Robertson effect
for irradiated particle with spherically symmetric mass distribution.
This yields linear relation between the incoming and outgoing momenta
(per unit time) of the radiation, in the proper reference frame of the particle:
$\vec{p}'_{o}$ $=$ ( 1 $-$ $\bar{Q}'_{pr, 1}$ $/$ $\bar{Q}'_{ext}$ )
$\vec{p}'_{i}$, where $\bar{Q}'_{pr, 1}$ and $\bar{Q}'_{ext}$ are
dimensionless efficiency factors for the radial direction radiation pressure
and extinction, integrated over the stellar spectrum (see Eq. 73,
or Eq. 74 for covariant formulation). A simple condition
0 $<$ $\bar{Q}'_{pr, 1}$ $/$ $\bar{Q}'_{ext}$ $\le$ 2 is obtained from the
invariant mass of the outgoing radiation (see Eqs. 77-79).
The case of "perfectly absorbing" spherical dust particle corresponds to
$\bar{Q}'_{pr, 1}$ $/$ $\bar{Q}'_{ext}$ = 1/2, if radius of the particle
is much larger than the wavelength(s) of the incident radiation: the
effect of diffraction cannot be neglected within geometrical optics
approximation. While the condition $\vec{p}'_{o}$
$=$ ( 1 $-$ $\bar{Q}'_{pr, 1}$ $/$ $\bar{Q}'_{ext}$ ) $\vec{p}'_{i}$
is consistent with the condition
$d M_{out}$ $=$ $\sqrt{d p_{out ~\mu} ~d p_{out}^{\mu}}$ $/$ $c^{2}$,
these fundamental relations are not consistent with the explanations
of the Poynting-Robertson effect in the literature.

The Poynting-Robertson effect is generated by simultaneous
action of the Doppler effect, change of concentration of photons
(the corresponding terms $w_{1}$ $\times$ $w_{1}$ are present at flux
density of radiation energy $S$ in equation of motion of spherical
particle, in Eq. 70) and the aberration of light, together with the laws
of relativity theory. However, the term $-$ $\vec{v} / c$
does not correspond to the aberration of light. It's existence is caused by
conservation of mass of the particle.
The term $1 / w_{1}$ in the four-vector $b_{1}^{\mu}$ (see Eq. 70)
comes from the Doppler effect (see Eqs. 17-18) or from the change of
concentration of photons (see Eqs. 21-22): Lorentz transformation
of the zero-th component of a four-vector is relevant, in relation
between $\vec{e}'_{1}$ and $\vec{e}_{1}$.

Covariant four-accelerations for the incoming and outgoing radiation
are given by Eqs. (106) and (109).
Moreover, the outgoing radiation accelerates the particle if
$\bar{Q}'_{pr, 1} / \bar{Q}'_{ext}$ $<$ 1 (see Eq. 110), i. e., also the
perfectly absorbing sphere treated by Poynting (1903), Robertson (1937)
and others.

As for the incoming radiation, the formulae for the energy and the mass of the
radiation are given in Eq. (50). The statement that the mass corresponds
to the energy through the relation "mass equals energy $/$ $c^{2}$" is not
correct, because the photon is massless (see also Sec. 8).
The relation between the incoming energy and the
increase of the particle's mass is presented in Eq. (62) (Eq. 63 presents the
result for momentum). As a consequence, the formula
''energy equals $\gamma$ $\times$ mass $\times$ $c^{2}$'' does not hold.
The Einstein's principle of equivalence of inertial mass and rest-energy
holds (Einstein 1999, p. 43).

As for the outgoing radiation: if the particle is under the action of incident
radiation, then the formulae for the energy and mass of the
radiation are given by Eq. (54), see also Eq. (55): the simple formula
''energy equals $\gamma$ times mass times $c^{2}$'' holds (see also Eq. 55).
The relation between the outgoing energy and
decrease of the particle's mass is presented in Eq. (66) for general case,
and in Eq. (89) for particle with spherically symmetric mass distribution.
As a consequence, the formula ''energy equals
$\gamma$ $\times$ mass $\times$ $c^{2}$'' does not hold, in general.
The Einstein's principle of equivalence of inertial mass and rest-energy
holds.

Eqs. (40)-(41) present effect of thermal emission alone, if the particle
is not irradiated. As a consequence, the isotropic thermal emission does not
influence acceleration of the particle. Eq. (41) (Eq. 67) also
shows that the formulation ''energy equals mass times $c^{2}$'' holds
between the decrease of particle's mass and the thermally emitted energy
in the rest frame of the particle. General equation is given by Eq. (68)
and the thermal force disturbs the saying
''energy equals $\gamma$ $\times$ mass $\times$ $c^{2}$''.
The relation between energy and mass of the radiation is given by Eqs.
(58) and (59): the standard simple formula
''energy equals $\gamma$ $\times$ mass $\times$ $c^{2}$'' holds.

On the basis of the covariant formulations presented in the paper, we
were able to show that three of the current statements on the essence of the
Poynting-Robertson effect are not correct. We have explained the
important points in the statements.

The relativistically covariant formulations
presented in the paper enable to understand: (i) physics of the
Poynting-Robertson effect (see also Secs. 7.3.4 and 7.4),
which is frequently used in studies of orbital
evolution of cosmic dust particles, and, (ii) why the analogy on momentum
transfer to the surface of an object in mechanics and electromagnetism
does not hold (see Sec. 7.5).

We have also shown an application of four-momentum (and mass)
of the outgoing radiation for arbitrarily shaped dust particle on
obtaining some information on cross sections, unknown from experiments,
observations or theoretical/numerical solutions. The condition is
presented in Eq. (158). As a result, the non-radial components of radiation
pressure cross sections decrease the ratio $\bar{C}'_{pr, ~1}$ $/$
$\bar{C}'_{ext}$ in comparison with the spherical particles. A more general
condition is given by Eq. (156). The condition can help in better
understanding of optical properties of cosmic dust particles (special cases
of the condition, applied to the spherical and planar particles, enabled us
to understand physics of the interaction between the incident
electromagnetic radiation and the particles, see Eqs. 79, 104,
Secs. 7.3.4, 7.4, 7.5).

The physics presented in the paper has direct implications
for understanding of distribution and evolution of cosmic dust grains in
various astrophysical systems.

\begin{acknowledgements}
This work was supported by the Scientific Grant Agency VEGA, Slovakia,
grant No. 1/3074/06.
\end{acknowledgements}


\begin{thebibliography}{}
\bibitem{}Beiser A., 1969. Perspectives of Modern Physics,
MvGraw-Hill Books, New York.
\bibitem{}Bernstein J., Fishbane P. M., Gasiorowicz S., 2000. Modern Physics,
Prentice Hall, Upper Saddle River, 602 pp.
\bibitem{}Bohren C. F., Huffman D. R., 1983. Absorption and Scattering of
Light by Small Particles, John Wiley $\&$ Sons, Inc., New York.
\bibitem{}Burns J. A., Lamy P. L., Soter S., 1979. Radiation forces on small
particles in the Solar System. {\it Icarus} {\bf 40}, 1-48.
\bibitem{}Carroll B. W., Ostlie D. A., 2007. An Introduction to Modern
Astrophysics, Pearson Education, Inc., Publishing as Addison-Wesley,
San Francisco, 2nd ed., 1278 pp.
\bibitem{}Chow T. L., 2008. Gravity, Black Holes, and the Very Early Universe
Springer Science+Business Media, LLC, New York, 280 pp.
\bibitem{}Danby J. M. A., 2003. Fundamentals of Celestial Mechanics.
Willmann-Bell, Inc., Richmond, 2nd ed., 483 pp.
\bibitem{}Demtr\H{o}der W., 2006. Atoms, Molecules and Photons: An Introduction
to Atomic-, Molecular- and Quantum-Physics, Springer-Verlag, Berlin, 571 pp.
\bibitem{}Dermott S. F., Jayraman S., Xu Y. L., Gustafson B. A. S.,  Liou J. C.,
1994. A circumsolar ring of asteroidal dust in resonant lock with the
Earth. {\it Nature} {\bf 369}, 719-723.
\bibitem{}Dohnanyi J. S., 1978. Particle dynamics. In: Cosmic Dust,
J. A. M. McDonnell (Ed.), Wiley-Interscience, Chichester, 527-605.
\bibitem{}Einstein A., 1905. Zur Elektrodynamik der bewegter K\H{o}rper.
{\it Annalen der Physik} {\bf 17}, 891-920.
\bibitem{}Einstein A., 1906. The principle of conservation of motion of the
center of gravity and the inertia of energy.
{\it Annalen der Physik} {\bf 20}, 627-633.
\bibitem{}Einstein A., 1999. Elementary derivation of the equivalence of mass
and energy. {\it Bull. (New Series) Amer. Math. Soc.} {\bf 37}, 39-44.
(Reprinted from {\it Bull. (New Series) Amer. Math. Soc.} {\bf 41} (1935),
223-230.)
\bibitem{}Ferraro R., 2007. Einstein's Space-Time: An Introduction to
Special and General Relativity, Springer Science $+$ Business Media, LLC,
New York, 310 pp.
\bibitem{}Festou M. C., Keller H. U., Weaver H. A. (Eds.), 2004.
Comets II. The University of Arizona Press, Tucson, in collaboration with
Lunar and Planetary Institute, Houston.
\bibitem{}Gasiorowicz S., 1974. Quantum Physics,
John Wiley $\&$ Sons, Inc., New York, 514 pp.
\bibitem{}Gustafson, B. A. S., 1994. Physics of Zodiacal Dust.
{\it Annual Review of Earth and Planetary Sciences} {\bf 22}, 553-595.
\bibitem{}Gr\H{u}n E., 2007. Solar System dust.
In: Encyclopedia of the Solar System, L.-A. McFadden, P. R.
Weissmann and T. V. Johnson (eds.), Academic Press (Elsevier), San Diego,
2nd ed., 621-636.
\bibitem{}Halliday D., Resnick R., Walker J., 2008. Fundamentals of Physics,
John Wiley $\&$ Sons, Inc., Hoboken, 8th ed., 1248 pp.
\bibitem{}Harwit M., 2006. Astrophysical Concepts.
Springer, New York, 4th ed., 714 pp.
\bibitem{}Jackson A. A., Zook H. A., 1989. A Solar System dust ring with the
Earth as its shepherd. {\it Nature} {\bf 337}, 629-631.
\bibitem{}Jewett J. W., Serway R. A., 2008. Physics for Scientists and
Engineers with Modern Physics, Thomson Learning, Inc., London, 7th ed., 1392 pp.
\bibitem{}Kapi\v{s}insk\'{y} I., 1984. Nongravitational effects affecting small
meteoroids in interplanetary space. {\it Contr. Astron. Obs. Skalnat\'{e}
Pleso} {\bf 12}, 99-111.
\bibitem{}Karttunen H., Kr\H{o}ger P., Oja H., Poutanen M., Donner K. J. (Eds.),
2007. Fundamental Astronomy, Springer-Verlag, Berlin Heidelberg,
5th ed., 510 pp.
\bibitem{}Kittel Ch., Knight W. D., Ruderman M. A., 1962. Mechanics. Berkeley
Physics Course - Volume 1, McGraw-Hill Book Company, New York, 480 pp.
\bibitem{}Kla\v{c}ka J., 1992. Poynting-Robertson effect. I. Equation of
motion. {\it Earth, Moon, and Planets} {\bf 59}, 41-59.
\bibitem{}Kla\v{c}ka J., 1993. Interplanetary dust particles: disintegration
and orbital motion. {\it Earth, Moon, and Planets} {\bf 60}, 17-21.
\bibitem{}Kla\v{c}ka J., 1994. Interplanetary dust particles and solar
radiation. {\it Earth, Moon, and Planets} {\bf 64}, 125-132.
\bibitem{}Kla\v{c}ka J., 2000. Electromagnetic radiation and motion of real
particle. \\
http://xxx.lanl.gov/abs/astro-ph/0008510
\bibitem{}Kla\v{c}ka J., 2004. Electromagnetic radiation and motion of a
particle, {\it Celestial Mech. and Dynam. Astron.} {\bf 89}, 1-61.
\bibitem{}Kla\v{c}ka J., Kocifaj M., 1994. Electromagnetic radiation
and equation of motion for a dust particle. In: Dynamics and Astrometry
of Natural and Artificial Celestial Bodies, K. Kurzynska, F. Barlier,
P. K. Seidelmann and I. Wytrzyszczak (Eds.), Astronomical Observatory
of A. Mickiewicz Observatory, Poznan, 187-190.
\bibitem{}Kla\v{c}ka J., Kocifaj M., 2001. Motion of nonspherical dust
particle under the action of electromagnetic radiation,
{\it J. Quant. Spectrosc. Radiat. Transfer} {\bf 70}, 595-610.
{\it Astron. Astrophys.} {\bf 464}, 127-134.
\bibitem{}Kocifaj M., Kla\v{c}ka J., Kundrac\'{\i}k F., 2000. Motion of
realistically shaped cosmic dust particle in Solar System. In:
Light Scattering by Nonspherical Particles: Halifax Contributions,
G. Videen, Q. Fu and P. Ch\'{y}lek (Eds.), Army Research Laboratory,
Adelphi, Maryland, 257-261.
\bibitem{}Krauss O., Wurm G., 2004. Radiation pressure forces on individual
micron-size dust particles: a new experimental approach,
{\it J. Quant. Spectrosc. Radiat. Transfer} {\bf 89}, 179-189.
\bibitem{}Kr\"{u}gel E., 2008. An introduction to the physics of
interstellar dust. Taylor \& Francis, Boca Raton, 387pp.
\bibitem{}Landau L. D., Lifshitz E. M., 2005. The Classical Theory of Fields,
Course of Theoretical Physics, Volume 2, Elsevier Butterworth-Heinemann,
Oxford, 4th ed., 428 pp.
\bibitem{}Leinert Ch., Gr\H{u}n E., 1990. Interplanetary dust. In:
Physics of the Inner Heliosphere I,
R. Schwen and E. Marsch (Eds.), Spriger-Verlag, Berlin, 207-275.
\bibitem{}Lissauer J. J., Murray C. D., 2007. Solar System Dynamics: Regular and
Chaotic Motion. In: Encyclopedia of the Solar System, L-A. McFadden,
P. R. Weissmann and T. V. Johnson (Eds.), Academic Press (Elsevier),
San Diego, 2nd ed., 787-812.
\bibitem{}Mie G., 1908. Beitr\H{a}ge zur Optik tr\H{u}ber Medien speziell
kolloidaler Metal\H{o}sungen. {\it Ann. Phys.} {\bf 25}, 377-445.
\bibitem{}Mishchenko M., 2001. Radiation force caused by scattering, absorption,
and emission of light by nonspherical particles.
{\it J. Quant. Spectrosc. Radiat. Transfer} {\bf 70}, 811-816.
\bibitem{}Mishchenko M., Travis L. D., Lacis A. A., 2002. Scattering, Absorption
and Emission of Light by Small Particles. Cambridge University Press,
Cambridge (UK), 445 pp.
\bibitem{}Murray C. D., Dermott S. F., 1999. Solar System Dynamics.
Cambridge University Press, Cambridge.
\bibitem{}Okun L. B., 1989a. The concept of mass (mass, energy, relativity).
{\it Sov. Phys. Usp.} {\bf 32}, 629-638.
\bibitem{}Okun L. B., 1989b. The concept of mass.
{\it Physics Today}, June 1989, 31-36.
\bibitem{}Poynting J. M., 1903. Radiation in the Solar System: its Effect on
Temperature and its Pressure on Small Bodies. {\it Philosophical Transactions
of the Royal Society of London} {\bf Series A 202}, 525-552.
\bibitem{}Quinn, T., 2005. Planet Formation.
In: Chaos and Stability in Planetary Systems,
Dvorak R., Freistetter F., Kurths J. (eds.), Springer-Verlag, Berlin, 187-217.
\bibitem{}Reach W. T., Franz B. A., Welland J. L., Hauser M. G., Kelsall T. N.,
Wright E. L., Rawley G., Stemwedel S. W., Splesman W. J., 1995.
Observational confirmation of a circumsolar dust ring by the COBE
satellite. {\it Nature} {\bf 374}, 521-523.
\bibitem{}Robertson H. P., 1937. Dynamical effects of radiation in the Solar
System. {\it Mon. Not. R. Astron. Soc.} {\bf 97}, 423-438.
\bibitem{}Robertson H. P., Noonan T. W., 1968. Relativity and Cosmology.
Saunders, Philadelphia, 456 pp.
\bibitem{}Sachs M., 2007. Concepts of Modern Physics: The Haifa Lectures.
Imperial College Press, London, 129 pp.
\bibitem{}Serway R. A., Moses C. J., Moyer C. A., 2005. Modern Physics,
Brooks/Cole - Thomson Learning, Belmont, 3rd ed., 600 pp.
\bibitem{}Schr\"{o}der U. E., 1990. Special Relativity, World Scientific
Publishing, Co. Pte. Ltd., Singapore, 214 pp.
\bibitem{}Sykes M. V., 2007. Infrared Views of the Solar System from
Space. In: Encyclopedia of the Solar System, L.-A. McFadden, P. R.
Weissmann and T. V. Johnson (eds.), Academic Press (Elsevier), San Diego,
2nd ed., 681-694.
\bibitem{}Taylor E. F., Wheeler J. A., 1992. Spacetime Physics:
Introduction to Special Relativity, W. H. Freeman and Company, New York,
2nd ed., 312 pp.
\bibitem{}Ugarov V. A., 1969. Special Theory of Relativity, Nauka, Moscow,
304 pp. (in Russian)
\bibitem{}van de Hulst H. C., 1981. Light Scattering by Small Particles.
Dover Publications, Inc. New York, 470 pp.
(originally published in 1957 by John Wiley \& Sons, Inc., New York)
\bibitem{}Woolfson M., 2000. The Origin and Evolution of the Solar System.
Institute of Physics Publishing, Bristol, 420 pp.
\bibitem{}Wyatt S. P., Whipple F. L., 1950. The Poynting-Robertson effect on
meteor orbits. {\it Astrophys. J.} {\bf 111}, 558-565.
\end{thebibliography}
\end{document}